\documentstyle[12pt]{article}
\begin{document}
\baselineskip=24pt
\def\rd{{\rm d}}
\newcommand{\Lamb}{\Lambda_{b}}
\newcommand{\Lamc}{\Lambda_{c}}
\newcommand{\omeb}{\omega_{b}^{(*)}}
\newcommand{\omec}{\omega_{c}^{(*)}}
\newcommand{\omeq}{\omega_{Q}^{(*)}}
\newcommand{\dsp}{\displaystyle}
\newcommand{\nn}{\nonumber}
\newcommand{\dfr}[2]{ \displaystyle\frac{#1}{#2} }
\newcommand{\Lag}{\Lambda \scriptscriptstyle _{ \rm GR} } 
\newcommand{\pa}{p\parallel}
\newcommand{\pe}{p\perp}
\newcommand{\pet}{p\top}
\newcommand{\paa}{p'\parallel}
\newcommand{\pee}{p'\perp} 
\newcommand{\pete}{p'\top} 
\renewcommand{\baselinestretch}{1.5}
\begin{titlepage}
\vspace{-20ex}
\vspace{1cm}
%\begin{flushright}
%\vspace{-3.0ex} 
%    {\sf ADP-98-14/T290} \\
%\vspace{-2.0mm}    
%       \it{}\\
%\vspace{-2.0mm}
%\vspace{5.0ex}
%\end{flushright}

\centerline{\Large\sf Bethe-Salpeter 
Equation for Heavy Baryons $\omega_{Q}^{(*)}$
in the Diquark Picture}
\vspace{6.4ex}
\centerline{\large\sf  	
X.-H. Guo$^{1,2}$, A.W. Thomas$^{1}$  and  A.G. Williams$^{1}$}
\vspace{3.5ex}
\centerline{\sf $^1$ Department of Physics and Mathematical Physics,}
\centerline{\sf and Special Research Center for the Subatomic Structure of
Matter,}
\centerline{\sf University of Adelaide, SA 5005, Australia}
\centerline{\sf $^2$ Institute of High Energy Physics, Academia Sinica,
Beijing 100039, China}
\centerline{\sf e-mail:  xhguo@physics.adelaide.edu.au,
athomas@physics.adelaide.edu.au,}
\centerline{\sf awilliam@physics.adelaide.edu.au}
\vspace{6ex}
\begin{center}
\begin{minipage}{5in}
\centerline{\large\sf 	Abstract}
\vspace{1.5ex}
\small {In the heavy quark limit, the heavy baryons $\omeq$ ($\omega$ 
stands for
$\Sigma$, $\Xi$ or $\Omega$ and $Q=b$ or $c$) are regarded as composed of a 
heavy quark and an axial vector, light diquark with good spin and isospin 
quantum numbers. Based on this diquark picture we establish the 
Bethe-Salpeter (B-S) equation for $\omeq$ in the limit
where the heavy quark has infinite mass,
$m_Q \rightarrow \infty$. It is found that in this limit there are three
components in the B-S wave function for $\omeq$.
Assuming the kernel to consist of a scalar confinement term and
a one-gluon-exchange term we derive three coupled integral equations for the
three B-S scalar functions in the covariant instantaneous approximation. 
Numerical solutions for the three B-S scalar functions are presented,
including a discussion of their dependence on the various input parameters.
These solutions are applied to calculate the Isgur-Wise functions 
$\xi (\omega)$ and $\zeta (\omega)$ for the weak transitions 
$\Omega_{b}^{(*)} \rightarrow \Omega_{c}^{(*)}$. Using these
%Isgur-Wise functions 
we give predictions for the Cabibbo-allowed
nonleptonic decay widths and
up-down asymmetries for $\Omega_{b} \rightarrow \Omega_{c}^{(*)}$ plus a
pseudoscalar or vector meson.}

\end{minipage}
\end{center}

\vspace{1cm}

\end{titlepage}
\vspace{0.2in}
{\large\bf I. Introduction}
\vspace{0.2in}

The physics of heavy hadrons has been a subject of intense interest 
in recent years. One reason for this is that more and 
more experimental data are being accumulated.
Another reason is the discovery of the new flavor and spin symmetry in QCD,
$SU(2)_f \times SU(2)_s$, in the heavy quark limit and the establishment 
of heavy quark effective theory (HQET) \cite{wise}. 
However, in comparison with the heavy meson case, 
heavy baryons have been studied much less, both
experimentally and theoretically. 

On the other hand, 
the experimental situation with heavy baryons has been improving 
recently, with more measurements becoming available.
For instance, OPAL has measured some physical
quantities for $\Lambda_b$, such as its lifetime and the production 
branching ratio for the inclusive semileptonic decay 
$\Lambda_b \rightarrow \Lambda l^- \bar{\nu} X$ \cite{opal}. Furthermore, 
measurements of the nonleptonic decay of $\Lambda_b$ have also been made,
through the well-known process $\Lambda_b \rightarrow \Lambda J/\psi$. 
The discrepancy between the measurements made by UA1 \cite{ua1} and 
those by CDF \cite{cdf2} and LEP
\cite{lep} appears to have  been settled by the new measurement from
CDF \cite{cdf1}. However, compared with D and B mesons, the
data for heavy baryons are still very limited. Besides
$\Lamb$, there has been few data on other bottom baryons \cite{particle},
although we expect 
more data to appear in the near future. Clearly the time is right for
serious theoretical studies
of heavy baryon properties to begin.

Theoretically, the HQET can simplify the physical processes involving 
heavy quarks, since with the aid of the HQET
the number of independent form factors is reduced. For instance, in 
leading order of the $1/m_Q$ expansion only one form factor (the
Isgur-Wise function) remains for the $\Lamb \rightarrow \Lamc$
transition, while for $\omeb \rightarrow \omec$ (we follow the notations 
of Ref.\cite{mannel}, $\omega$ could be
$\Xi$, $\Sigma$ or $\Omega$ and $Q=b$ or $c$) there are two independent
Isgur-Wise functions. (Note that $\omega_{Q}^{(*)}$ is a notation
implying either $\omega_{Q}$ or $\omega_{Q}^{*}$.)
The behaviour of these functions 
depends on the nonperturbative effects of QCD which control the dynamics
inside a heavy hadron. Hence some nonperturbative QCD model has to be
adopted from which these Isgur-Wise functions can be obtained. 
In previous work \cite{bsguo}, we established the B-S equation
for $\Lambda_Q$, which is assumed to be composed of a heavy quark, Q, and
a scalar diquark. Some theoretical predictions for $\Lamb \rightarrow \Lamc$
were also obtained. It is the purpose of the present paper to generalize
such an approach to the heavy baryons, $\omeq$, and consequently give 
some phenomenological predictions for the weak decays of such baryons.

When the quark mass is very heavy compared with the QCD scale,
$\Lambda_{\rm QCD}$, the light degrees of freedom in a heavy baryon  
$\Lambda_{Q}$ ($Q=b$ or $c$) become blind to the flavor and spin quantum
numbers of the heavy quark because of the $SU(2)_f \times 
SU(2)_s$ symmetry. Therefore, 
the light degrees  of freedom have good quantum 
numbers, including angular momentum and isospin. These quantum numbers
can be used to classify heavy baryons. For example, 
the light degrees of freedom of $\Lambda_Q$  have zero
angular momentum and isospin.
For $\Sigma_{Q}^{(*)}$, the angular momentum and parity $J^P$ of the light
degrees of freedom are
$1^+$, and the isospin is also 1 in order to guarantee that
the total wave function
of the hadron is antisymmetric. Hence it is natural 
to consider the heavy baryon to be composed of a heavy quark and a light 
diquark. This is our underlying assumption. 
%Actually the diquark picture
%in heavy baryon system has been used in some other works\cite{guo}.

Based on the picture of the composition of the heavy baryon which we
have just presented, the
three body system is simplified to a two body system. We establish the 
B-S equation for the heavy baryons, $\omeq$, in this picture. The heavy
quark symmetry can be used to simplify the form of the B-S wave function
greatly. It can be shown that in the limit $m_Q \rightarrow \infty$
there are three components in the B-S wave function, and hence we have three
corresponding scalar functions.
We solve the B-S equation numerically by assuming that 
the kernel contains a scalar confinement term and
a one-gluon-exchange term. The explicit dependence of the B-S wave function
on the parameters of the model will be discussed. Furthermore, we calculate the
Isgur-Wise functions in terms of the B-S wave functions and give theoretical
predictions for the Cabbibo-allowed two body nonleptonic decays $\Omega_{b}
\rightarrow \Omega_{c}^{(*)}$ plus a pseudoscalar or vector meson. 

The light degrees of freedom of $\omeq$ belong to a 6 representation of
flavor SU(3). Taking $Q=b$ as an example,  $\omeb$ includes 
$\Sigma_{b}^{(*) +,0,-}$, $\Xi_{b}^{(*) 0,-}$ and $\Omega_{b}^{(*) -}$.
The total spin of $\omega_Q$ and $\omega_{Q}^{*}$ are $\frac{1}{2}$ and
$\frac{3}{2}$ respectively. There is no strange quark in $\Sigma_{Q}^{(*)}$
while there is one strange quark in $\Xi_{Q}^{(*)}$ and two in
$\Omega_{Q}^{(*)}$ respectively.    

The remainder of this paper is organized as follows. In Section II we
establish the B-S equation for the heavy quark and axial vector light
diquark system and discuss the form of the kernel. In Section III we derive 
explicitly the coupled integral equations for the B-S scalar wave functions.
In Section IV we discuss the normalization condition of the B-S wave 
function by exploiting the normalization of the Isgur-Wise function 
at the zero recoil
point. The numerical solutions of the B-S equation and their dependence
on the parameters in our model are presented in Section V.
In Section VI we calculate the Isgur-Wise functions and give predictions
for the decay widths and
up-down asymmetry parameters for  $\Omega_{b} \rightarrow \Omega_{c}^{(*)}$
plus a pseudoscalar or vector meson. 
Finally, Section VI contains a summary and discussions.

\vspace{0.2in}
{\large\bf II. The B-S equation for $\omeq$}
\vspace{0.2in}

As discussed in Section I, $\omeq$ is regarded as a bound state
of a heavy quark, $\psi_Q$, and a light axial vector diquark $A_{\mu}$. 
In the following, $u_Q$ denotes the Dirac spinor of $\psi_Q$ and
$\eta_{\mu}$ represents the polarization vector of $A_\mu$. 
Then $B_\mu\equiv u_Q \eta_{\mu}$ can be decomposed into spin-$\frac{1}{2}$ and
spin-$\frac{3}{2}$ states \cite{falk,mannel2} which represent
$\omega_Q$ and $\omega_{Q}^{*}$ respectively. These two states are
degenerate in the heavy quark limit. Using the notation of 
Refs. \cite{mannel2,guo2}
this doublet is described by $B_{\mu}^{m}(v)$, where $m=1,2$ correspond to
$\omega_Q$ and $\omega_{Q}^{*}$ respectively, $v_\mu$ is the velocity
of the heavy baryon and $B_\mu=B_{\mu}^{1}+B_{\mu}^{2}$ \cite{falk}.
Explicitly, we can write
\begin{eqnarray}
B_{\mu}^{1}(v)&=&\frac{1}{\sqrt{3}}(\gamma_\mu +v_\mu)\gamma_5 u(v), \nn \\
B_{\mu}^{2}(v)&=&u_\mu (v),
\label{2a}
\vspace{2mm}
\end{eqnarray}
where $u(v)$ is the Dirac spinor and $u_\mu (v)$ is the Rarita-Schwinger 
vector spinor. 
$B_{\mu}^{m}(v)$ satisfies the following conditions:
\begin{equation}
\rlap / vB_{\mu}^{m}(v)=B_{\mu}^{m}(v), \;\;\; v^\mu B_{\mu}^{m}(v)=0, 
\;\;\; \gamma^\mu B_{\mu}^{2}(v)=0.
\label{2b}
\vspace{2mm}
\end{equation}
The above constraints for $m=1$ can be seen from $\rlap / v u(v)=u(v)$ while
for $m=2$, they are the properties of a spin-$\frac{3}{2}$,
Rarita-Schwinger vector spinor.
% $B_{\mu}^{2}(v)$.

Under a Lorentz transformation $\Lambda$\cite{falk}
\begin{equation}
B_{\mu}\rightarrow \Lambda^{\nu}_{\mu}D(\Lambda)B_{\nu}, 
\label{2c}
\vspace{2mm}
\end{equation}
where $D(\Lambda)$ is the spinorial representation of $\Lambda$. Under
the heavy quark spin transformation\cite{mannel}
\begin{equation}
B_{\mu}\rightarrow -\gamma_5 \rlap /v \rlap /e B_{\mu}, 
\label{2d}
\vspace{2mm}
\end{equation}
where $e=e_1, e_2, e_3$ are three mutually orthogonal four-dimensional
unit vectors which are also 
orthogonal to $v$, (i.e., $e_i \cdot v=0$), and $e_{i}^{2}=-1\; (i=1,2,3)$. 
The three unit vectors are associated with the heavy quark
spin operators  which are generators of the $SU(2)_s$ symmetry.

Since $\omeq$ is composed of $\psi_Q$ and $A_\mu$,
we can define the B-S wave function of $\omeq$ by
\begin{equation}
\chi_{\mu} (x_1, x_2, P)=<0|T \psi_Q(x_1) A_\mu(x_2)|\omeq (P)>,
\label{2e}
\vspace{2mm}
\end{equation}
where $P=m_{\omeq}v$ is the total momentum of $\omeq$ and $v$ is its velocity.
Let $m_Q$ and $m_D$  be the masses of the heavy quark and the light diquark 
in the baryon. Let us define $\lambda_1\equiv\frac{m_Q}{m_Q+m_D}$ and
$\lambda_2\equiv\frac{m_D}{m_Q+m_D}$ and let
$p$ be the relative momentum of the two constituents. The B-S wave function in
momentum space is defined as
\begin{equation}
\chi_{\mu}^{m}(x_1, x_2, P)=e^{iP\cdot X}\int \frac{\rd^4 p}{(2\pi)^4}
e^{ip\cdot x}\chi_{P\mu}^{m}(p),
\label{2f}
\vspace{2mm}
\end{equation}
where $X\equiv\lambda_1 x_1+\lambda_2 x_2$ is the coordinate of the center of 
mass and $x\equiv x_1-x_2$. The momentum of the heavy quark is
$p_1=\lambda_1 P+p$ and that of the diquark is $p_2=-\lambda_2 P+p$.

It can be shown that $\chi_{Pm}^{\mu}(p)$ satisfies the following 
B-S equation\cite{lurie}
\begin{equation}
\chi_{Pm}^{\mu}(p)=S_F(\lambda_1 P+p)\int \frac{\rd^4q}{(2\pi)^4}
G_{\rho\nu}(P,p,q)\chi_{Pm}^{\nu}(q)
S_{D}^{\mu\rho}(-\lambda_2 P+p),
\label{2g}
\vspace{2mm}
\end{equation}
where $G_{\rho\nu}(P,p,q)$ is the kernel, which is defined as the sum of 
all the two particle irreducible diagrams with respect to the heavy quark and
the light diquark. For convenience, in the following we use the variables 
\begin{equation}
p_l\equiv v\cdot p-\lambda_2 m_{\omeq},\;\;\; p_t\equiv p-(v\cdot p)v. 
\label{2g1}
\vspace{2mm}
\end{equation}
Then in the 
leading order of the $1/m_Q$ expansion we have
\begin{equation}
S_F(\lambda_1 P+p)=\frac{i(1+\rlap/v)}{2(p_l+E_0+m_D+i\epsilon)}, 
\label{2h}
\vspace{2mm}
\end{equation}
where $E_0$ is the binding energy. From Eqs.(\ref{2g}) and 
(\ref{2h}) it follows
that $\chi_{Pm}^{\mu}(p)$ satisfies the following equation
\begin{equation}
\rlap /v\chi_{Pm}^{\mu}(p)=\chi_{Pm}^{\mu}(p).
\label{2hh}
\vspace{2mm}
\end{equation}
The propagator of the light axial vector diquark has the form
\begin{equation}
S_{D\mu\nu}(p_2)=-i\frac{g_{\mu\nu}-p_{2\mu}p_{2\nu}/m_{D}^{2}}
{p_{2}^{2}-m_{D}^{2}+i\epsilon}. 
\label{2i}
\vspace{2mm}
\end{equation}
In the limit $m_Q \rightarrow \infty$ we have $p_{2}=-m_D v +p$ and 
hence we have
\begin{equation}
S_{D\mu\nu}(p_2)=-i\;\frac{g_{\mu\nu}-v_\mu v_\nu-p_{\mu}p_{\nu}/m_{D}^{2}
+(v_\mu p_\nu +v_\nu p_\mu)/m_D}{p_{l}^{2}-W_{p}^{2}+i\epsilon}, 
\label{2j}
\vspace{2mm}
\end{equation}
where $W_{p}\equiv \sqrt{p_{t}^{2}+m_{D}^{2}}$.
The corrections to Eqs.(\ref{2h}) and (\ref{2j}) are from $O(1/m_Q)$ terms.

Now we discuss the form of the B-S wave function $\chi_{Pm}^{\mu}(p)$.
In the heavy quark limit, due to the $SU(2)_s\times SU(2)_f$ symmetry, the
internal dynamics of the heavy baryon, $\omeq$, is determined by the light
degrees of freedom and the flavor and spin direction
of the heavy quark, $Q$, is irrelevant. Consequently we have\cite{falk, guo2}
\begin{equation}
\chi_{P}^{\mu}(p)=u_Q(v) \eta_\nu  \zeta^{\mu\nu}(v,p),
\label{2k}
\vspace{2mm}
\end{equation}
and 
\begin{equation}
<0|A_\mu|{\rm light}, 1^+>=\eta_\nu  \zeta^{\mu\nu}(v,p).
\label{2l}
\vspace{2mm}
\end{equation}

Since $v\cdot \eta =0$\cite{falk}, 
the tensor $\zeta^{\mu\nu}(v,p)$ can be expanded as 
\begin{equation}
\zeta^{\mu\nu}(v,p)=A_1 g^{\mu\nu}+A_2 v^{\mu}p^{\nu}+A_3 p^{\mu}p^{\nu},
\label{2m}
\vspace{2mm}
\end{equation}
where $A_i (i=1,2,3)$ are Lorentz scalar functions. After expressing 
$u_Q \eta_\mu$ in Eq.(\ref{2k}) in terms 
of $B_{\mu}^{m}(v)\; (m=1,2)$,
we have the following form for the B-S
wave function 
\begin{equation}
\chi_{Pm}^{\mu}=A_1 B_{m}^{\mu}(v)+A_2 v^{\mu}p_{\nu} B_{m}^{\nu}(v)+
A_3 p^{\mu}p_{\nu} B_{m}^{\nu}(v).
\label{2n}
\vspace{2mm}
\end{equation}
 
Therefore, we have three components in the B-S wave function,
$\chi_{Pm}^{\mu}(p)$, and they correspond to three scalar B-S functions
$A_i (i=1,2,3)$. This is consistent with our diquark picture for 
$\omeq$. In the heavy quark limit, the dynamics inside the heavy
baryon is controlled by the configuration of the light degrees of freedom.
Since the light diquark is a $1^+$ object, it has three different
configurations. Consequently there are three components in the B-S wave
function which describe the dynamics in the heavy baryon $\omeq$.

In fact, we can derive the form of $\chi_{Pm}^{\mu}(p)$ in another way.
We may first write out all the possible terms which have the same behaviour
as $\chi_{Pm}^{\mu}(p)$ under Lorentz transformations. Then by applying
the condition Eq.(\ref{2hh}) and ensuring 
the proper behaviour under the heavy quark spin
transformation, Eq.(\ref{2d}), we obtain the same result as given in
Eq.(\ref{2n}).

Considering $p^\mu=v\cdot p v^{\mu}+p_{t}^{\mu}$, 
and using the constraint $v^\mu B_{\mu}^{m}(v)=0$,
it will be convenient to define 
$$A=A_1, \;\; C=A_2+v\cdot p A_3, \;\; D=A_3,$$
which results in the following expression for the B-S wave function:
\begin{equation}
\chi_{Pm}^{\mu}=A B_{m}^{\mu}(v)+C v^{\mu}p_{t\nu} B_{m}^{\nu}(v)+
D p_{t}^{\mu}p_{t\nu} B_{m}^{\nu}(v).
\label{2o}
\vspace{2mm}
\end{equation}

$A, C$ and $D$ in Eq.(\ref{2o}) are functions of $p_l$ and $p_{t}^{2}$. Their 
behaviour is controlled by nonperturbative QCD. Our aim is to
obtain explicit forms for them with some QCD-motivated model for the 
form of the B-S kernel. 
%Therefore, we
%have to use some nonperturbative QCD model for the kernel.

Motivated by the success
of the potential model\cite{eichten}, scalar confinement
and one-gluon-exchange terms were used 
in the kernel when studying $\Lambda_Q$
in Ref.\cite{bsguo}. This form was also used in the heavy meson case in 
Ref.\cite{dai}. In the present work we will also adopt this form
of the kernel
\begin{equation}
iG^{\rho\nu}=g^{\rho\nu}I\otimes I V_1 +v_{\mu} \otimes \Gamma^{\mu\rho\nu} 
V_2, 
\label{2p}
\vspace{2mm}
\end{equation}
where $\Gamma^{\mu\rho\nu}$ is the vertex of a gluon with two axial vector
diquarks. This vertex should reflect the internal structure of the diquark.
In this work, we
use the model proposed in Ref.\cite{kroll} where this vertex has the
following form (see Fig. 1)
\begin{equation}
-i\frac{\lambda^a}{2}g_s \Gamma^{\mu\rho\nu}F_V(Q^2), 
\label{2q}
\vspace{2mm}
\end{equation}
with 
$$\Gamma^{\mu\rho\nu}=(p_2+p'_2)^\mu g^{\nu\rho}-(p_{2}^{\nu}g^{\mu\rho}
+p_{2}^{\prime\rho}g^{\mu\nu}).$$
In Eq.(\ref{2q}) $g_s$ is the strong interaction coupling constant and 
$F_V(Q^2)$ is introduced to describe the internal structure of the axial
vector diquark. The form factor, $F_V(Q^2)$,
depends on nonperturbative QCD interactions and
will be determined phenomenologically, by comparison with experiment. 

As discussed in Ref.\cite{bsguo}, when we
consider the vertex of two heavy quarks with a gluon, the momenta of 
the two heavy quarks are $p_1=\lambda_1 m_{\Lambda_Q}v+p$ and
$p'_1=\lambda_1 m_{\Lambda_Q}v+q$ respectively, where $p$ 
and $q$ are relative momenta and of the order $\Lambda_{QCD}$. 
In the heavy quark limit the heavy quark is almost on-shell and moves with 
constant velocity. It can be shown that $p_l=q_l$ at this
vertex when the heavy quark is exactly 
on-shell. This is the so-called covariant instantaneous approximation
\cite{bsguo, dai}. With this approximation, $V_1$ and
$V_2$ in $G^{\rho\nu}(P,p,q)$ are replaced by 
\begin{equation}
\tilde{V}_i\equiv V_i|_{p_l=q_l} (i=1,2).
\label{2r}
\vspace{2mm}
\end{equation}

\vspace{0.2in}
{\large\bf III. Coupled integral equations for three B-S scalar wave functions}
\vspace{0.2in}

In this section we will derive explicitly three coupled integral equations 
for the B-S scalar wave functions.
Substituting Eqs.(\ref{2h}) and (\ref{2j}) into Eq.(\ref{2g}) 
and considering the form of the kernel in Eq.(\ref{2p}) and the property 
in Eq.(\ref{2hh}),
we obtain the following form for the B-S equation 
\begin{equation}
\chi_{Pm}^{\mu}(p)=\frac{-i}{(p_l+E_0+m_D+i\epsilon)(p_{l}^{2}-W_{p}^{2}
+i\epsilon)}M_{m}^{\mu},
\label{3a}
\vspace{2mm}
\end{equation}
where
\begin{equation}
M_{m\mu}\equiv i\left[g_{\mu\rho}-v_\mu v_\rho 
+\frac{(v_\mu p_\rho +p_\mu v_\rho)}
{m_D}-\frac{p_\mu p_\rho}{m_{D}^{2}}\right]
\int \frac{\rd^4 q}{(2\pi)^4}[G^{\rho\nu}(P,p,q)\chi_{Pm\nu}(q)]\mid_{p_l=q_l},
\label{3b}
\vspace{2mm}
\end{equation}
and we have made explicit use of the covariant instantaneous approximation.

Substituting Eqs.(\ref{2o}), (\ref{2p}) and 
(\ref{2q}) into Eq.(\ref{3b}) and again using
the covariant instantaneous approximation we have
\begin{eqnarray}
M_{m}^{\mu} &=& B_{m}^{\mu}(v)\int \frac{\rd^4 q}{(2\pi)^4}A(\tilde{V}_1+2p_l
\tilde{V}_2) +\frac{1}{m_D}v^\mu\int\frac{\rd^4 q}{(2\pi)^4}\{p_t\cdot B_m(v)
A(\tilde{V}_1+2p_l\tilde{V}_2) \nn \\
&& +(p_l+m_D)[-A p_t \cdot B_m(v) \tilde{V}_2+q_t\cdot B_m(v) (C\tilde{V}_1
-Dp_t \cdot q_t \tilde{V}_2)]\nn \\
&& +p_t \cdot q_t q_t\cdot B_m(v) [-C \tilde{V}_2+D(\tilde{V}_1+2p_l
\tilde{V}_2)]\} -\frac{1}{m_{D}^{2}}p^\mu\int\frac{\rd^4 q}{(2\pi)^4}\nn\\
&&\{p_t\cdot B_m(v)
A(\tilde{V}_1+2p_l\tilde{V}_2)
+p_l[-A p_t \cdot B_m(v) \tilde{V}_2 \nn\\
&&+q_t\cdot B_m(v) (C\tilde{V}_1
-Dp_t \cdot q_t \tilde{V}_2)]
+p_t \cdot q_t q_t\cdot B_m(v) [-C \tilde{V}_2+D(\tilde{V}_1+2p_l
\tilde{V}_2)]\} \nn \\
&&+\int \frac{\rd^4 q}{(2\pi)^4}q_{t}^{\mu}
q_t\cdot B_m(v) [-C \tilde{V}_2+D(\tilde{V}_1+2p_l
\tilde{V}_2)].
\label{3c}
\vspace{2mm}
\end{eqnarray}

We notice that in Eq.(\ref{3c}) there are terms of the form
$\int \frac{\rd^4 q}{(2\pi)^4}
q_{t}^{\mu}f$ and $\int \frac{\rd^4 q}{(2\pi)^4}q_{t}^{\mu}q_{t}^{\nu}f$,
where $f$ is some function of $p^2, q^2$, and $p\cdot q$. On the grounds 
of Lorentz invariance, in general we have
\begin{equation}
\int \frac{\rd^4 q}{(2\pi)^4}q_{t}^{\mu}f=f_1v^\mu+f_2p_{t}^{\mu},
\label{3d}
\vspace{2mm}
\end{equation}
and 
\begin{equation}
\int \frac{\rd^4 q}{(2\pi)^4}q_{t}^{\mu}q_{t}^{\nu}f=g_1g^{\mu\nu}+
g_2v^\mu v^\nu+g_3 v^\mu p_{t}^{\nu}+g_4v^\nu p_{t}^{\mu}+g_5p_{t}^{\mu} 
p_{t}^{\nu}.
\label{3e}
\vspace{2mm}
\end{equation}
From Eq.(\ref{3d}) only the $f_2$ term can contribute, 
while from Eq.(\ref{3e}) 
only the $g_1, g_3$ and $g_5$ terms can contribute, 
since $v_\nu B^{\nu}_{m}(v)=0$.
It can be easily shown that
\begin{eqnarray}
f_2&=&\int \frac{\rd^4 q}{(2\pi)^4}\frac{p_t\cdot q_t}{p_{t}^{2}}f,\nn\\
g_1&=&\int \frac{\rd^4 q}{(2\pi)^4}\frac{(p_t\cdot q_t)^2-p_{t}^{2}
q_{t}^{2}}{2p_{t}^{2}}f,\nn\\
g_3&=&0, \nn\\
g_5&=&\int \frac{\rd^4 q}{(2\pi)^4}\frac{3(p_t\cdot q_t)^2-p_{t}^{2}
q_{t}^{2}}{2p_{t}^{4}}f.
\label{3ee}
\vspace{2mm}
\end{eqnarray}

With the aid of Eqs.(\ref{3d}), (\ref{3e}) and (\ref{3ee}) 
we can express $M_{m}^{\mu}$ in Eq.(\ref{3c})
in terms of  $B_{m}^{\mu}(v)$, $v^{\mu}p_{t\nu} B_{m}^{\nu}(v)$ and
$p_{t}^{\mu}p_{t\nu} B_{m}^{\nu}(v)$. Let us define 
\begin{equation}
\tilde{A}(p_{t}^{2})=\int \frac{\rd q_l}{2\pi}A(p_l,p_{t}^{2}),\;\;\;\;
\tilde{C}(p_{t}^{2})=\int \frac{\rd q_l}
{2\pi}C(p_l,p_{t}^{2}),\;\;\;\;\tilde{D}(p_{t}^{2})
=\int \frac{\rd q_l}{2\pi}D(p_l,p_{t}^{2}),
\label{3f}
\vspace{2mm}
\end{equation}
where $\tilde{A}$, $\tilde{C}$ and $\tilde{D}$ are functions of $p_{t}^{2}$
only. Then one obtains the expression
\begin{eqnarray}
M_{m}^{\mu} &=& B_{m}^{\mu}(v)\int \frac{\rd^3 q_t}{(2\pi)^3}\left\{\tilde{A}
(\tilde{V}_1+2p_l\tilde{V}_2) -\tilde{C}\frac{(p_t\cdot q_t)^2-p_{t}^{2}
q_{t}^{2}}{2p_{t}^{2}}\tilde{V}_2 \right.\nn\\
&&\left.+\tilde{D}\frac{(p_t\cdot q_t)^2-p_{t}^{2}
q_{t}^{2}}{2p_{t}^{2}}(\tilde{V}_1+2p_l\tilde{V}_2)\right\} \nn\\
&&+\frac{1}{m_{D}^{2}}v^\mu p_t\cdot B_m(v)
\int\frac{\rd^3 q_t}{(2\pi)^3}\left\{-\tilde{A}
[p_l\tilde{V}_1+(p_{l}^{2}+m_{D}^{2})\tilde{V}_2] \right. \nn\\
&&\left.-\tilde{C}\left[(p_{l}^{2}-m_{D}^{2})\frac{p_t\cdot q_t}{p_{t}^{2}}
\tilde{V}_1+p_l \frac{(p_t\cdot q_t)^2}{p_{t}^{2}}\tilde{V}_2\right] \right.
\nn\\
&&\left.+\tilde{D}\frac{(p_t\cdot q_t)^2}{p_{t}^{2}}[p_{l}
\tilde{V}_1+(p_{l}^{2}+m_{D}^{2})\tilde{V}_2]\right\} \nn\\
&&-\frac{1}{m_{D}^{2}}p_{t}^{\mu} p_t\cdot B_m(v)
\int\frac{\rd^3 q_t}{(2\pi)^3}\left\{\tilde{A}
(\tilde{V}_1+p_{l}\tilde{V}_2) \right.\nn\\
&&\left.+\tilde{C}\left[p_{l}\frac{p_t\cdot q_t}{p_{t}^{2}}
\tilde{V}_1+\frac{m_{D}^{2}(3(p_t\cdot q_t)^2-p_{t}^{2}
q_{t}^{2})+2p_{t}^{2}(p_t\cdot q_t)^2}{2p_{t}^{4}}\tilde{V}_2\right]\right. 
\nn\\
&&\left.+\tilde{D}\left[-\frac{m_{D}^{2}(3(p_t\cdot q_t)^2-p_{t}^{2}
q_{t}^{2})+2p_{t}^{2}(p_t\cdot q_t)^2}{2p_{t}^{4}}
(\tilde{V}_1+2p_l\tilde{V}_2) \right.\right.\nn\\
&&\left.\left.+p_l \frac{(p_t\cdot q_t)^2}{p_{t}^{2}}\tilde{V}_2\right]
\right\}.
\label{3g}
\vspace{2mm}
\end{eqnarray}

In Eq.(\ref{3a}) there are poles in $p_l$ at 
$W_p-i\epsilon$,$-W_p+i\epsilon$ and
$-E_0-m_D-i\epsilon$. By choosing the appropriate contour, 
we integrate over $p_l$ on both sides of Eq.(\ref{3a}) and 
obtain the following, three coupled integral equations for 
$\tilde{A}$,$\tilde{C}$ and $\tilde{D}$
\begin{eqnarray}
\tilde{A}(p_{t}^{2})&=&\frac{-1}{2W_p(E_0+m_D-W_p)}\int \frac{\rd^3 q_t}
{(2\pi)^3}\left\{\tilde{A}(q_{t}^{2})(\tilde{V}_1-2W_p\tilde{V}_2)\right.\nn\\
&&\left.-\tilde{C}(q_{t}^{2})\frac{(p_t\cdot q_t)^2-p_{t}^{2}
q_{t}^{2}}{2p_{t}^{2}}\tilde{V}_2 
+\tilde{D}(q_{t}^{2})\frac{(p_t\cdot q_t)^2-p_{t}^{2}
q_{t}^{2}}{2p_{t}^{2}}(\tilde{V}_1-2W_p\tilde{V}_2)\right\}, \nn\\
&& 
\label{3h}
\vspace{2mm}
\end{eqnarray}
\begin{eqnarray}
\tilde{C}(p_{t}^{2})&=&\frac{-1}{2m_{D}^{2}W_p(E_0+m_D-W_p)}
\int \frac{\rd^3 q_t}
{(2\pi)^3}\left\{\tilde{A}(q_{t}^{2})[W_p\tilde{V}_1 \right. \nn\\
&&\left.
-((E_0+m_D)W_p+m_{D}^{2})\tilde{V}_2]\right.\nn\\
&&\left.+\tilde{C}(q_{t}^{2})\left[-\frac{p_t\cdot q_t}{p_{t}^{2}}
((E_0+m_D)W_p-m_{D}^{2})\tilde{V}_1+W_p\frac{(p_t\cdot q_t)^2}{p_{t}^{2}} 
\tilde{V}_2\right]\right. \nn\\
&&\left.+\tilde{D}(q_{t}^{2})\frac{(p_t\cdot q_t)^2}{p_{t}^{2}}
[-W_p\tilde{V}_1+((E_0+m_D)W_p+m_{D}^{2})
\tilde{V}_2]\right\},
\label{3i}
\vspace{2mm}
\end{eqnarray}
\begin{eqnarray}
\tilde{D}(p_{t}^{2})&=&\frac{1}{2m_{D}^{2}W_p(E_0+m_D-W_p)}
\int \frac{\rd^3 q_t}
{(2\pi)^3}\left\{\tilde{A}(q_{t}^{2})(\tilde{V}_1-W_p\tilde{V}_2)\right.\nn\\
&&\left.+\tilde{C}(q_{t}^{2})\left[-\frac{p_t\cdot q_t}{p_{t}^{2}}
W_p\tilde{V}_1+\frac{m_{D}^{2}(3(p_t\cdot q_t)^2-p_{t}^{2}
q_{t}^{2})+2p_{t}^{2}(p_t\cdot q_t)^2}{2p_{t}^{4}}
\tilde{V}_2\right]\right. \nn\\
&&\left.-\tilde{D}(q_{t}^{2})\left[\frac{m_{D}^{2}(3(p_t\cdot q_t)^2-p_{t}^{2}
q_{t}^{2})+2p_{t}^{2}(p_t\cdot q_t)^2}{2p_{t}^{4}}
(\tilde{V}_1-2W_p\tilde{V}_2)\right.\right.\nn\\
&&\left.\left.+\frac{(p_t\cdot q_t)^2}{p_{t}^{2}}
W_p\tilde{V}_2\right]\right\}.
\label{3j}
\vspace{2mm}
\end{eqnarray}

If one knows the form for the kernel, $\tilde{V}_1$ and $\tilde{V}_2$, 
then $\tilde{A(p_{t}^{2})}$,$\tilde{C}(p_{t}^{2})$ and $\tilde{D}(p_{t}^{2})$
can be obtained from Eqs.(\ref{3h}), (\ref{3i}) and (\ref{3j}). 
Consequently from Eqs. (\ref{2o}), (\ref{3a}) and (\ref{3g}) 
we find the following expressions
for $A(p_l,p_{t}^{2})$, $C(p_l,p_{t}^{2})$ and $D(p_l,p_{t}^{2})$: 
\begin{eqnarray}
A(p_l,p_{t}^{2})&=&\frac{-i}{(p_l+E_0+m_D+i\epsilon)(p_{l}^{2}-W_{p}^{2}
+i\epsilon)}\int \frac{\rd^3 q_t}
{(2\pi)^3}\left\{\tilde{A}(q_{t}^{2})(\tilde{V}_1+2p_l\tilde{V}_2)\right.\nn\\
&&\left.-\tilde{C}(q_{t}^{2})\frac{(p_t\cdot q_t)^2-p_{t}^{2}
q_{t}^{2}}{2p_{t}^{2}}\tilde{V}_2 
+\tilde{D}(q_{t}^{2})\frac{(p_t\cdot q_t)^2-p_{t}^{2}
q_{t}^{2}}{2p_{t}^{2}}(\tilde{V}_1+2p_l\tilde{V}_2)\right\},\nn\\
&& 
\label{3k}
\vspace{2mm}
\end{eqnarray}
\begin{eqnarray}
C(p_l,p_{t}^{2})&=&\frac{-i}{m_{D}^{2}(p_l+E_0+m_D+i\epsilon)
(p_{l}^{2}-W_{p}^{2}+i\epsilon)}
\int \frac{\rd^3 q_t}
{(2\pi)^3} 
\left\{-\tilde{A}(q_{t}^{2})[p_l\tilde{V}_1 \right.\nn\\ 
&&\left.+(p_{l}^{2}+m_{D}^{2})\tilde{V}_2]
-\tilde{C}(q_{t}^{2})\left[(p_{l}^{2}-m_{D}^{2})
\frac{p_t\cdot q_t}{p_{t}^{2}}
\tilde{V}_1+p_l\frac{(p_t\cdot q_t)^2}{p_{t}^{2}} 
\tilde{V}_2\right]\right. \nn\\
&&\left.+\tilde{D}(q_{t}^{2})\frac{(p_t\cdot q_t)^2}{p_{t}^{2}}
[p_l\tilde{V}_1+(p_{l}^{2}+m_{D}^{2})
\tilde{V}_2]\right\},
\label{3l}
\vspace{2mm}
\end{eqnarray}
\begin{eqnarray}
D(p_l,p_{t}^{2})&=&\frac{i}{m_{D}^{2}(p_l+E_0+m_D+i\epsilon)
(p_{l}^{2}-W_{p}^{2}+i\epsilon)}
\int \frac{\rd^3 q_t}
{(2\pi)^3}\left\{\tilde{A}(q_{t}^{2})(\tilde{V}_1+p_l\tilde{V}_2)\right.\nn\\
&&\left.+\tilde{C}(q_{t}^{2})\left[\frac{p_t\cdot q_t}{p_{t}^{2}}
p_l\tilde{V}_1+\frac{m_{D}^{2}(3(p_t\cdot q_t)^2-p_{t}^{2}
q_{t}^{2})+2p_{t}^{2}(p_t\cdot q_t)^2}{2p_{t}^{4}}
\tilde{V}_2\right]\right. \nn\\
&&\left.+\tilde{D}(q_{t}^{2})\left[-\frac{m_{D}^{2}(3(p_t\cdot q_t)^2-p_{t}^{2}
q_{t}^{2})+2p_{t}^{2}(p_t\cdot q_t)^2}{2p_{t}^{4}}
(\tilde{V}_1+2p_l\tilde{V}_2)\right.\right.\nn\\
&&\left.\left.+\frac{(p_t\cdot q_t)^2}{p_{t}^{2}}
p_l\tilde{V}_2\right]\right\}.
\label{3m}
\vspace{2mm}
\end{eqnarray}

A model kernel, specified in terms of
$\tilde{V}_1$ and $\tilde{V}_2$, for the B-S equation in the
scalar light diquark case was given in Ref.\cite{bsguo}. In the present 
axial vector
diquark case, a model vertex for a gluon with two $1^+$ diquarks is given in
Eq.(\ref{2q}), where $F_V(Q^2)$ describes the internal structure of the
light diquark. Following Ref.\cite{kroll} we take the form for $F_V(Q^2)$
as
\begin{equation}
F_V(Q^2)=\frac{\alpha_{s}^{({\rm eff})}Q_{1}^{2}}{Q^{2}+Q_{1}^{2}},
\label{3n}
\vspace{2mm}
\end{equation}
where $Q_{1}^{2}$ is a parameter which freezes $F_V(Q^2)$ when $Q^2$ is
very small. In the high energy region the form factor is proportional
to $1/Q^2$, which is consistent with perturbative QCD calculations
\cite{brodsky}. By analyzing the electromagnetic form factor
for the proton it was found that selecting $Q_{1}^{2}=3.2$GeV$^2$ can lead to
results consistent with the experimental data \cite{kroll}. Note 
that in Eq.(\ref{3n}) we do not consider the difference between longitudinal
and transverse polarization states. The reason is that we are considering the
bound state with a binding energy of order $\Lambda_{\rm QCD}$, 
so this difference
should be small (see Ref.\cite{kroll}, where this difference is a factor
$\frac{Q_{2}^{2}}{Q^{2}+Q_{2}^{2}}$ with $Q_{2}^{2}$
approximately 15GeV$^2$, so this factor is close to 1 in our discussions).

Based on the above discussion, the kernel for the B-S equation in the
baryon case is taken to have the following form
\begin{eqnarray}
\tilde{V}_1&=&\frac{8\pi\kappa}{[(p_t-q_t)^2+\mu^2]^2}-(2\pi)^3
\delta^3  (p_t-q_t)
	\int \frac{\rd^3 k}{(2\pi)^3}\frac{8\pi\kappa}{(k^2+\mu^2)^2}, \nn \\
\tilde{V}_2&=&-\frac{16\pi}{3}
	\frac{\alpha_{s}^{({\rm eff})2}
Q_{1}^{2}}{[(p_t-q_t)^2+\mu^2][(p_t-q_t)^2+Q_{1}^{2}]},
\label{3o}
\vspace{2mm}
\end{eqnarray}
where $\kappa$ and $\alpha_{s}^{({\rm eff})}$ are coupling parameters related
to scalar confinement and the one-gluon-exchange diagram
respectively. The second term in Eq.(\ref{3o}) is the counter term
which removes the infra-red divergence arising from the linear confinement
in the integral equation.
The parameter $\mu$ is introduced to avoid the infra-red divergence in 
numerical calculations. The limit $\mu \rightarrow 0$ is taken in the end.
Besides $Q_{1}^{2}$, there are two parameters $\kappa$, and 
$\alpha_{s}^{({\rm eff})}$, in the kernel. However, they should be related 
to each other when we solve the coupled integral equations (\ref{3h}),
(\ref{3i}) and (\ref{3j}).
This will be discussed in detail in Section V.

Since we now have an explicit form for $\tilde{V}_1$ and $\tilde{V}_2$ in 
Eq.(\ref{3o}), we can reduce Eqs.(\ref{3h}), (\ref{3i}) and
(\ref{3j}) to one dimensional integral equations.
With the aid of the formulas given in Appendix A we obtain the following
equations from Eqs.(\ref{3h}), (\ref{3i}) and (\ref{3j}).
\begin{eqnarray}
\tilde{A}(p_{t}^{2})&=&\frac{-1}{2W_p(E_0+m_D-W_p)}\int 
\frac{q_{t}^{2}\rd q_t}{4\pi^2}
\{[8\pi\kappa F_1(|p_{t}|,|q_{t}|)+\frac{32\pi\beta W_p}{3
(Q_{1}^{2}-\mu^2)} \nn\\
&&(F_2(|p_{t}|,|q_{t}|,\mu)-F_2(|p_{t}|,|q_{t}|,Q_1))]
\tilde{A}(q_{t}^{2})-8\pi\kappa F_1(|p_{t}|,|q_{t}|)\tilde{A}(p_{t}^{2})\}\nn\\
&&-\frac{1}{2W_p(E_0+m_D-W_p)}\int \frac{q_{t}^{2}\rd q_t}{4\pi^2}
\frac{8\pi\beta}{3(Q_{1}^{2}-\mu^2)}\{q_{t}^{2}
[-F_2(|p_{t}|,|q_{t}|,\mu)\nn\\
&&+F_2(|p_{t}|,|q_{t}|,Q_1)]-\frac{1}{p_{t}^{2}}
[(F_4(|p_{t}|,|q_{t}|,\mu)-F_4(|p_{t}|,|q_{t}|,Q_1)]\}
\tilde{C}(q_{t}^{2})\nn\\
&&+\frac{1}{2W_p(E_0+m_D-W_p)}\int \frac{q_{t}^{2}\rd q_t}{4\pi^2}
\{4\pi\kappa [q_{t}^{2}F_1(|p_{t}|,|q_{t}|)-\frac{1}{p_{t}^{2}}
F_5(|p_{t}|,|q_{t}|)]\nn\\
&&+\frac{16\pi\beta W_p}{3(Q_{1}^{2}-\mu^2)}[q_{t}^{2}
(F_2(|p_{t}|,|q_{t}|,\mu)-F_2(|p_{t}|,|q_{t}|,Q_1))\nn\\
&&+\frac{1}{2p_{t}^{2}}(F_4(|p_{t}|,|q_{t}|,\mu)-F_4(|p_{t}|,|q_{t}|,Q_1))]\}
\tilde{D}(q_{t}^{2}),
\label{3p}
\vspace{2mm}
\end{eqnarray}
\begin{eqnarray}
\tilde{C}(p_{t}^{2})&=&\frac{-1}{2m_{D}^{2}W_p(E_0+m_D-W_p)}\int 
\frac{q_{t}^{2}\rd q_t}{4\pi^2}
\{[8\pi\kappa W_p F_1(|p_{t}|,|q_{t}|)+\frac{16\pi\beta}{3
(Q_{1}^{2}-\mu^2)} \nn\\
&&(F_2(|p_{t}|,|q_{t}|,\mu)-F_2(|p_{t}|,|q_{t}|,Q_1))
((E_0+m_D)W_p+m_{D}^{2})]
\tilde{A}(q_{t}^{2})\nn\\
&&-8\pi\kappa W_p 
F_1(|p_{t}|,|q_{t}|)\tilde{A}(p_{t}^{2})\}
-\frac{1}{2m_{D}^{2}W_p(E_0+m_D-W_p)}\int \frac{q_{t}^{2}\rd q_t}{4\pi^2}\nn\\
&&
\{[8\pi\kappa F_3(|p_{t}|,|q_{t}|)\frac{m_{D}^{2}-(E_0+m_D)W_p}{p_{t}^{2}}\nn\\
&&+\frac{16\pi\beta}{3(Q_{1}^{2}-\mu^2)}
(F_4(|p_{t}|,|q_{t}|,\mu)-F_4(|p_{t}|,|q_{t}|,Q_1))\frac{W_p}{p_{t}^{2}}]
\tilde{C}(q_{t}^{2})\nn\\
&&-8\pi\kappa F_1(|p_{t}|,|q_{t}|)[m_{D}^{2}-(E_0+m_D)W_p]
\tilde{C}(p_{t}^{2})\}\nn\\
&&+\frac{1}{2m_{D}^{2}W_p(E_0+m_D-W_p)}\int \frac{q_{t}^{2}\rd q_t}{4\pi^2}
\{[8\pi\kappa F_5(|p_{t}|,|q_{t}|)\frac{W_p}{p_{t}^{2}} \nn\\
&&-\frac{16\pi\beta}{3(Q_{1}^{2}-\mu^2)}
(F_4(|p_{t}|,|q_{t}|,\mu)-F_4(|p_{t}|,|q_{t}|,Q_1))\nn\\
&&\frac{(E_0+m_D)
W_p +m_{D}^{2}}{p_{t}^{2}}]\tilde{D}(q_{t}^{2})
-8\pi\kappa p_{t}^{2} W_p F_1(|p_{t}|,|q_{t}|)\tilde{D}(p_{t}^{2})\},
\label{3q}
\vspace{2mm}
\end{eqnarray}
\begin{eqnarray}
\tilde{D}(p_{t}^{2})&=&\frac{1}{2m_{D}^{2}W_p(E_0+m_D-W_p)}\int 
\frac{q_{t}^{2}\rd q_t}{4\pi^2}
\{[8\pi\kappa F_1(|p_{t}|,|q_{t}|)+\frac{16\pi\beta W_p}{3
(Q_{1}^{2}-\mu^2)} \nn\\
&&(F_2(|p_{t}|,|q_{t}|,\mu)-F_2(|p_{t}|,|q_{t}|,Q_1))]
\tilde{A}(q_{t}^{2})-8\pi\kappa  
F_1(|p_{t}|,|q_{t}|)\tilde{A}(p_{t}^{2})\}\nn\\
&&+\frac{1}{2m_{D}^{2}W_p(E_0+m_D-W_p)}\int\frac{q_{t}^{2}\rd q_t}{4\pi^2}
\{[-8\pi\kappa F_3(|p_{t}|,|q_{t}|)\frac{W_p}{p_{t}^{2}}\nn\\
&&+\frac{16\pi\beta}{3(Q_{1}^{2}-\mu^2)}
(F_4(|p_{t}|,|q_{t}|,\mu)-F_4(|p_{t}|,|q_{t}|,Q_1))\frac{2p_{t}^{2}
+3m_{D}^{2}}{2p_{t}^{4}} \nn\\
&&+\frac{16\pi\beta}{3(Q_{1}^{2}-\mu^2)}
(F_2(|p_{t}|,|q_{t}|,\mu)-F_2(|p_{t}|,|q_{t}|,Q_1))\frac{q_{t}^{2}
m_{D}^{2}}{2p_{t}^{2}}]
\tilde{C}(q_{t}^{2})\nn\\
&&+8\pi\kappa W_p F_1(|p_{t}|,|q_{t}|)
\tilde{C}(p_{t}^{2})\}\nn\\
&&-\frac{1}{2m_{D}^{2}W_p(E_0+m_D-W_p)}\int \frac{q_{t}^{2}\rd q_t}{4\pi^2}
\{[8\pi\kappa F_5(|p_{t}|,|q_{t}|)\frac{2p_{t}^{2}+3m_{D}^{2}}
{2p_{t}^{4}} \nn\\
&&-8\pi\kappa F_1(|p_{t}|,|q_{t}|)\frac{q_{t}^{2}m_{D}^{2}}{2p_{t}^{2}}
-\frac{16\pi\beta}{3(Q_{1}^{2}-\mu^2)}\nn\\
&&(F_4(|p_{t}|,|q_{t}|,\mu)-F_4(|p_{t}|,|q_{t}|,Q_1))
\frac{W_p(p_{t}^{2}+3m_{D}^{2})}{p_{t}^{4}}\nn\\
&&-\frac{16\pi\beta}{3(Q_{1}^{2}-\mu^2)}(F_2(|p_{t}|,|q_{t}|,\mu)
-F_2(|p_{t}|,|q_{t}|,Q_1))
\frac{W_p m_{D}^{2} q_{t}^{2}}{p_{t}^{2}}
]\tilde{D}(q_{t}^{2})\nn\\
&&-8\pi\kappa (p_{t}^{2}+m_{D}^{2}) F_1(|p_{t}|,|q_{t}|)\tilde{D}(p_{t}^{2})\},
\label{3r}
\vspace{2mm}
\end{eqnarray}
where $|p_t|=\sqrt{p_{t}^{2}}$.
The functions $F_i (i=1,...,5)$, appearing in Eqs.(\ref{3p}), (\ref{3q}) 
and (\ref{3r}), are defined
in Appendix A. From these three coupled integral equations we can solve 
numerically for  
$\tilde{A}(p_{t}^{2})$, $\tilde{C}(p_{t}^{2})$ and $\tilde{D}(p_{t}^{2})$.
This will be done in Section V. In the next section, we will first
discuss the normalization of the B-S wave function.

\vspace{0.2in}
{\large\bf IV. Normalization for the B-S wave function}
\vspace{0.2in}

It can be seen that the overall normalization of 
$\tilde{A}(p_{t}^{2})$, $\tilde{C}(p_{t}^{2})$ and $\tilde{D}(p_{t}^{2})$
cannot be determined from Eqs.(\ref{3p}), (\ref{3q}) and
(\ref{3r}). With the help of heavy quark symmetry, the
normalization constant can be obtained from the fact that the Isgur-Wise
function is normalized to one at the zero-recoil point. In the limit
$m_Q \rightarrow \infty$ the weak transition matrix element induced
by the current $\bar{c}\Gamma b$ for
$\omeb\rightarrow\omec$ has the following form from the HQET,
\begin{equation}
\langle\omec (v')|\bar{c}\Gamma b|\omeb (v)\rangle 
=\bar{B}_{m'}^{\nu}(v')\Gamma
B_{m}^{\mu}(v)(\xi(\omega)g_{\mu\nu}+\zeta(\omega)v_\nu v'_\mu),
\label{4a}
\vspace{2mm}
\end{equation}
where $\omega=v\cdot v'$ is the velocity transfer, $m, m'$ could be 1 or 2,
and $\Gamma$ is an arbitrary Dirac matrix. At the zero-recoil point,
$v=v'$, only the $\xi(\omega)g_{\mu\nu}$ term contributes and we must have 
\begin{equation}
\xi(\omega=1)=1.
\label{4b}
\vspace{2mm}
\end{equation}

On the other hand, the transition matrix element for $\omeb \rightarrow
\omec$ is related to the B-S wave functions of $\omeb$ and $\omec$ by
the following equation
\begin{equation}
\langle \omec (v')|\bar{c}\Gamma b|\omeb (v)\rangle=\int
\frac{\rd^4p}{(2\pi)^4} \bar{\chi}_{P'm'}^{\mu}(p')\Gamma
\chi_{Pm}^{\nu}(p)S^{-1}_{D\mu\nu}(p_2),
\label{4bb}
\vspace{2mm}
\end{equation}
where $P$ ($P'$) is the momentum of $\omeb$ ($\omec$) and
$\bar{\chi}_{P'm'}^{\mu}(p')$ is the wave function of the final state $\omec
(v')$, which satisfies the constraint 
\begin{equation}
\bar{\chi}_{P'm'}^{\mu}(p')\rlap/v'=\bar{\chi}_{P'm'}^{\mu}(p').
\label{4c}
\vspace{2mm}
\end{equation}
At the zero-recoil point, $p'=p$, since the  light diquark sees no change
in the heavy quark part it does not change its relative
momentum. 

The scalar B-S functions of the final state B-S wave function obey the same
B-S equation as (\ref{3h}), (\ref{3i}) and
(\ref{3j}). Substituting Eq.(\ref{2g})
into Eq.(\ref{4bb}) and using Eq.(\ref{4a}) we have
\begin{equation}
\xi(1)\bar{B}_{m'\mu}(v)\Gamma B_{m}^{\mu}(v)=\int
\frac{\rd^4p}{(2\pi)^4}\frac{i}
{p_l+E_0+m_D+i\epsilon} \bar{\chi}_{Pm'}^{\mu}(p)\Gamma
\int\frac{\rd^4q}{(2\pi)^4}G_{\mu\nu}(P,p,q)
\chi_{Pm}^{\nu}(p).
\label{4d}
\vspace{2mm}
\end{equation}

Now we substitute the expression for the kernel Eq.(\ref{2p}) and Eq.(\ref{2o})
into Eq.(\ref{4d}). Using the same technique as used for 
Eqs.(\ref{3d}), (\ref{3e})
and (\ref{3ee}), we find that there is only the structure 
$\bar{B}_{m'}^{\mu}(v)\Gamma B_{m\mu}(v)$ on the right hand side of 
Eq.(\ref{4d}). Substituting the explicit expressions for $\tilde{V}_1$ and
$\tilde{V}_2$ in Eq.(\ref{3o}), and using the integration formula 
in Appendix A,
we arrive at the following
expression for $\xi(1)$ after some tedious calculations
\begin{eqnarray}
\xi(1)&=&\int\frac{p_{t}^{2}\rd p_t}{4\pi^2}\frac{2}{E_0+m_D-W_p}
\left[\tilde{A}(p_{t}^{2})h_1(|p_t|)-\frac{1}{3}
p_{t}^{2} \tilde{C}(p_{t}^{2})h_2(|p_t|)\right.\nn\\
&&\left.-\frac{1}{3}
p_{t}^{2} \tilde{D}(p_{t}^{2})h_3(|p_t|)+\frac{1}{6m_{D}^{2}}
p_{t}^{2} h_2(|p_t|)h_4(|p_t|)\right],
\label{4e}
\vspace{2mm}
\end{eqnarray}
where 
$h_i(|p_t|) (i=1,2,3,4)$ are given by the following equations 
\begin{eqnarray}
h_1(|p_t|)&=&\int \frac{q_{t}^{2}\rd q_t}{4\pi^2}
\{8\pi\kappa F_1(|p_{t}|,|q_{t}|)[\tilde{A}(q_{t}^{2})-\frac{1}{3}
q_{t}^{2}\tilde{D}(q_{t}^{2})]+\frac{16\pi\beta}{3
(Q_{1}^{2}-\mu^2)} \nn\\
&&[F_2(|p_{t}|,|q_{t}|,\mu)-F_2(|p_{t}|,|q_{t}|,Q_1)]
[2W_p\tilde{A}(q_{t}^{2})-\frac{1}{3}
q_{t}^{2}\tilde{C}(q_{t}^{2})\nn\\
&&-\frac{2}{3}W_p
q_{t}^{2}\tilde{D}(q_{t}^{2})]
-8\pi\kappa F_1(|p_{t}|,|q_{t}|)[\tilde{A}(p_{t}^{2})-\frac{1}{3}
p_{t}^{2}\tilde{D}(p_{t}^{2})]\},
\label{4f}
\vspace{2mm}
\end{eqnarray}
\begin{eqnarray}
h_2(|p_t|)&=&\int \frac{q_{t}^{2}\rd q_t}{4\pi^2}
\{\frac{16\pi\beta}{3(Q_{1}^{2}-\mu^2)} 
[F_2(|p_{t}|,|q_{t}|,\mu)-F_2(|p_{t}|,|q_{t}|,Q_1)]
\tilde{A}(q_{t}^{2})\nn\\
&&+8\pi\kappa F_3(|p_{t}|,|q_{t}|)\frac{1}{p_{t}^{2}}\tilde{C}(q_{t}^{2})
+\frac{16\pi\beta}{3(Q_{1}^{2}-\mu^2)} 
[F_4(|p_{t}|,|q_{t}|,\mu)\nn\\
&&-F_4(|p_{t}|,|q_{t}|,Q_1)]\frac{1}{p_{t}^{2}}
\tilde{D}(q_{t}^{2})-8\pi\kappa F_1(|p_{t}|,|q_{t}|)\tilde{C}(p_{t}^{2})\},
\label{4g}
\vspace{2mm}
\end{eqnarray}
\begin{eqnarray}
h_3(|p_t|)&=&\int \frac{q_{t}^{2}\rd q_t}{4\pi^2}
\{8\pi\kappa F_1(|p_{t}|,|q_{t}|)\tilde{A}(q_{t}^{2})
+\frac{32\pi\beta W_p}{3(Q_{1}^{2}-\mu^2)} 
[F_2(|p_{t}|,|q_{t}|,\mu)\nn\\
&&-F_2(|p_{t}|,|q_{t}|,Q_1)]
\tilde{A}(q_{t}^{2})-8\pi\kappa F_5(|p_{t}|,|q_{t}|)\frac{1}{p_{t}^{2}}
\tilde{D}(q_{t}^{2})
+\frac{16\pi\beta}{3(Q_{1}^{2}-\mu^2)} \nn\\
&&[F_4(|p_{t}|,|q_{t}|,\mu)-F_4(|p_{t}|,|q_{t}|,Q_1)]\frac{1}{p_{t}^{2}}
[\tilde{C}(q_{t}^{2})+2W_p\tilde{D}(q_{t}^{2})]\nn\\
&&-8\pi\kappa F_1(|p_{t}|,|q_{t}|)[-\tilde{A}(p_{t}^{2})+
p_{t}^{2}\tilde{D}(p_{t}^{2})]\},
\label{4h}
\vspace{2mm}
\end{eqnarray}
\begin{eqnarray}
h_4(|p_t|)&=&\int \frac{q_{t}^{2}\rd q_t}{4\pi^2}
\{\frac{-16\pi\beta}{3(Q_{1}^{2}-\mu^2)} 
[F_2(|p_{t}|,|q_{t}|,\mu)-F_2(|p_{t}|,|q_{t}|,Q_1)]
\tilde{A}(q_{t}^{2})\nn\\
&&+8\pi\kappa F_3(|p_{t}|,|q_{t}|)\frac{1}{p_{t}^{2}}\tilde{C}(q_{t}^{2})
-\frac{16\pi\beta}{3(Q_{1}^{2}-\mu^2)} 
[F_4(|p_{t}|,|q_{t}|,\mu)\nn\\
&&-F_4(|p_{t}|,|q_{t}|,Q_1)]\frac{1}{p_{t}^{2}}
\tilde{D}(q_{t}^{2})-8\pi\kappa F_1(|p_{t}|,|q_{t}|)\tilde{C}(p_{t}^{2})\}.
\label{4i}
\vspace{2mm}
\end{eqnarray}

The B-S scalar functions $\tilde{A}(p_{t}^{2})$, $\tilde{C}(p_{t}^{2})$
and $\tilde{D}(p_{t}^{2})$ should be normalized such that they satisfy 
Eq.(\ref{4e}).

%\vspace{2cm}
\vspace{0.2in}
{\large\bf V. Numerical solutions for the B-S wave function}
\vspace{0.2in}

In this section we solve the three coupled integral equations,  
Eqs.(\ref{3p}), (\ref{3q})
and (\ref{3r}), numerically. The method is to discretize the
integration region into $n$ pieces (with $n$ sufficiently large). 
In this way, the integral equations become matrix equations and
the B-S scalar functions $\tilde{A}(p_{t}^{2})$, $\tilde{C}(p_{t}^{2})$
and $\tilde{D}(p_{t}^{2})$ become $n$ dimensional vectors. The matrix
equations obtained in this way can be written in the following form,
\begin{equation}
\tilde{A}=Z_1 \tilde{A}+Z_2 \tilde{C}+Z_3 \tilde{D},
\label{5a}
\vspace{2mm}
\end{equation}
\begin{equation}
R_1 \tilde{A}+R_2 \tilde{C}+R_3 \tilde{D}=0,
\label{5b}
\vspace{2mm}
\end{equation}
\begin{equation}
T_1 \tilde{A}+T_2 \tilde{C}+T_3 \tilde{D}=0,
\label{5c}
\vspace{2mm}
\end{equation}
where $Z_i, R_i, T_i (i=1,2,3)$ are $n\times n$ matrices and are given by
Eqs. (\ref{3p}), (\ref{3q}) and (\ref{3r}).

Substituting Eqs.(\ref{5b}) and (\ref{5c}) into (\ref{5a}) we obtain the
eigenvalue equation for $\tilde{A}$
\begin{equation}
H\tilde{A}=\tilde{A},
\label{5d}
\vspace{2mm}
\end{equation}
where $H$ is an $n\times n$ matrix
\begin{equation}
H=Z_1+Z_2(T_{3}^{-1}T_2-R_{3}^{-1}R_2)^{-1}(R_{3}^{-1}R_1-T_{3}^{-1}T_1)
+Z_3(T_{2}^{-1}T_3-R_{2}^{-1}R_3)^{-1}(R_{2}^{-1}R_1-T_{2}^{-1}T_1).
\label{5e}
\vspace{2mm}
\end{equation}
The eigenvalue equation (\ref{5d}) is solved by the so-called inverse 
iteration method\cite{rae}. In this way, we first construct the operator
\begin{equation}
K=\frac{1}{H-\lambda},
\label{5f}
\vspace{2mm}
\end{equation}
where $\lambda$ is 
some parameter which is chosen to be near to the eigenvalue 1
in Eq.(\ref{5d}). In order to solve for the eigenvector $\tilde{A}$,
we start with an arbitrary vector $Y$ and operate $K$ on $Y$ sufficiently
many times so that the eigenvector corresponding to the eigenvalue 1
dominates. In this way, the scalar function $\tilde{A}$ is obtained.

In our model we have several parameters, $\alpha_{s}^{({\rm eff})}$, $\kappa$, 
$Q_{1}^{2}$, $m_D$ and $E_0$. The parameter  $Q_{1}^{2}$ has been described
in Section III, with $Q_{1}^{2}=3.2$GeV$^2$
from the data of the electromagnetic form factor of the proton. 
It is noted that this value corresponds to the $(qq')$ axial vector
diquark ($q, q' =u$ or $d$), i.e., for $\Sigma_{Q}^{(*)}$. In the cases of 
$\Xi_{Q}^{(*)}$ or $\Omega_{Q}^{(*)}$ this
value might be somewhat different 
because of $SU(3)$ flavor symmetry breaking. 
However, we do not have data to extract $Q_{1}^{2}$ for $\Xi_{Q}^{(*)}$
and $\Omega_{Q}^{(*)}$ at
present. In this work, we simply use the same value for $Q_{1}^{2}$
based on the approximate $SU(3)$ flavor symmetry. On the other hand,
the binding energy should satisfy the following relation
\begin{equation}
m_{\omeq}=m_Q+m_D+E_0,
\label{5g}
\vspace{2mm}
\end{equation}
where we have omitted corrections of $O(1/m_Q)$, since we are working 
in the heavy quark limit. Note that 
$m_D+E_0$ is independent of the flavor of the
heavy quark, because of the $SU(2)_f\times SU(2)_s$ symmetry.
From the B-S equation solutions in the meson
case, it has been found that the values $m_b=5.02$GeV and $m_c=1.58$GeV give
predictions which are in good agreement with experiments
\cite{dai}. Hence in the baryon case we expect
\begin{equation}
m_D+E_0=0.88GeV ({\rm for}\; \Sigma_{Q}^{(*)}),\;\;
1.07GeV ({\rm for} \;\Xi_{Q}^{(*)}),\;\; 1.12GeV ({\rm for} \;
\Omega_{Q}^{(*)}).
\label{5h}
\vspace{2mm}
\end{equation}

The parameter $m_D$ cannot be determined, although there are suggestions from 
the analysis of valence structure functions that it should be around 0.9GeV
for non-strange diquarks \cite{tony}. Hence we let it vary within some 
reasonable range.
When we solve the eigenvalue equation Eq.(\ref{5d}), the condition that
the eigenvalue is 1 provides a relation between $\alpha_{s \rm eff}$ 
and $\kappa$.
As discussed in Ref.\cite{bsguo} $\kappa$ is related to 
$\kappa'$ ($\kappa'$ is the confinement parameter in the heavy meson case
and is about 0.2GeV$^2$\cite{eichten, dai}), where  
$\kappa \sim \Lambda_{\rm QCD}\kappa'$.
Therefore, in our numerical calculations we let $\kappa$ vary in the
region between 0.02GeV$^3$ and 0.1GeV$^3$. The diquark mass, $m_D$, is
chosen to vary from 0.9GeV to 1GeV for $\Sigma_{Q}^{(*)}$,
from 1.1GeV to 1.2GeV for $\Xi_{Q}^{(*)}$, and from 1.15GeV to 1.25GeV for 
$\Omega_{Q}^{(*)}$. Then we obtain the parameter $\alpha_{s}^{({\rm eff})}$
for different values of $m_D$ and $\kappa$. 
The numerical results are shown in Tables 1,2 and 3 for $\Sigma_{Q}^{(*)}$,
$\Xi_{Q}^{(*)}$ and $\Omega_{Q}^{(*)}$, respectively.

\begin{table}
\caption{Values  of  $\kappa$  and  $\alpha_{s}^{({\rm eff})}$ 
for $\Sigma_{Q}^{(*)}$ with three sets of $m_D$}
\begin{center}
\begin{tabular}{lccccc}
\hline
\hline
$m_D$(GeV)& & &0.90  & &\\ 
\hline
$\kappa$(GeV$^3$)&0.02&0.04&0.06 &0.08 &0.10\\
\hline
$\alpha_{s}^{({\rm eff})}$ &0.5190&0.5593&0.5842&0.6061&0.6149\\
\hline
\hline
$m_D$(GeV)& & &0.95  & &\\ 
\hline
$\kappa$(GeV$^3$)&0.02&0.04&0.06 &0.08 &0.10\\
\hline
$\alpha_{s}^{({\rm eff})}$ &0.5889&0.6123&0.6285&0.6406&0.6502\\
\hline
\hline
$m_D$(GeV)& & &1.0  & &\\ 
\hline
$\kappa$(GeV$^3$)&0.02&0.04&0.06 &0.08 &0.10\\
\hline
$\alpha_{s}^{({\rm eff})}$ &0.6414&0.6560&0.6669&0.6757&0.6828\\
\hline
\hline
\end{tabular}
\end{center}
\end{table}

\begin{table}
\caption{Values  of  $\kappa$  and $\alpha_{s}^{({\rm eff})}$ 
for $\Xi_{Q}^{(*)}$ with three sets of $m_D$}
\begin{center}
\begin{tabular}{lccccc}
\hline
\hline
$m_D$(GeV)& & &1.10  & &\\ 
\hline
$\kappa$(GeV$^3$)&0.02&0.04&0.06 &0.08 &0.10\\
\hline
$\alpha_{s}^{({\rm eff})}$ &0.5047&0.5402&0.5643&0.5826&0.5974\\
\hline
\hline
$m_D$(GeV)& & &1.15  & &\\ 
\hline
$\kappa$(GeV$^3$)&0.02&0.04&0.06 &0.08 &0.10\\
\hline
$\alpha_{s}^{({\rm eff})}$ &0.5785&0.5995&0.6155&0.6283&0.6391\\
\hline
\hline
$m_D$(GeV)& & &1.20  & &\\ 
\hline
$\kappa$(GeV$^3$)&0.02&0.04&0.06 &0.08 &0.10\\
\hline
$\alpha_{s}^{({\rm eff})}$ &0.6341&0.6478&0.6588&0.6682&0.6763\\
\hline
\hline
\end{tabular}
\end{center}
\end{table}

\begin{table}
\caption{Values  of  $\kappa$  and  $\alpha_{s}^{({\rm eff})}$ 
for $\Omega_{Q}^{(*)}$ with three sets of $m_D$}
\begin{center}
\begin{tabular}{lccccc}
\hline
\hline
$m_D$(GeV)& & &1.15  & &\\ 
\hline
$\kappa$(GeV$^3$)&0.02&0.04&0.06 &0.08 &0.10\\
\hline
$\alpha_{s}^{({\rm eff})}$ &0.4975&0.5325&0.5565&0.5748&0.5897\\
\hline
\hline
$m_D$(GeV)& & &1.20  & &\\ 
\hline
$\kappa$(GeV$^3$)&0.02&0.04&0.06 &0.08 &0.10\\
\hline
$\alpha_{s}^{({\rm eff})}$ &0.5729&0.5935&0.6093&0.6222&0.6331\\
\hline
\hline
$m_D$(GeV)& & &1.25  & &\\ 
\hline
$\kappa$(GeV$^3$)&0.02&0.04&0.06 &0.08 &0.10\\
\hline
$\alpha_{s}^{({\rm eff})}$ &0.6296&0.6430&0.6539&0.6633&0.6714\\
\hline
\hline
\end{tabular}
\end{center}
\end{table}

With the parameters in Tables 1,2 and 3 we obtain the
numerical solution for the B-S scalar function $\tilde{A}$ as the eigenvector
of Eq.(\ref{5d}). Consequently, we get the numerical solutions for
$\tilde{C}$ and $\tilde{D}$ from Eqs.(\ref{5b}) and 
(\ref{5c}). These solutions 
depend on the parameters $m_D$ and $\kappa$.
In Figs. 2, 3 and 
4 we show the shapes of $\tilde{A}$, $\tilde{C}$ and $\tilde{D}$
for $\Sigma_{Q}^{(*)}$, $\Xi_{Q}^{(*)}$ and $\Omega_{Q}^{(*)}$ respectively.
Figs. 2(a), 3(a) and 4(a) show the dependence on $\kappa$ for a typical $m_D$,
while Figs. 2(b), 3(b) and 4(b)
show the dependence on $m_D$ for a typical $\kappa$.
It can be seen from these figures that for different heavy baryons the shape
of the B-S scalar functions are rather 
similar. This arises from the approximate $SU(3)$ flavor
symmetry and is to be expected. 
All the scalar functions decrease to zero when $|p_t|$ is larger than
about 1.5GeV, because of the 
confinement interaction. Furthermore, since we are discussing
the ground states $\omeq$, there are no nodes in the functions.

\vspace{0.2in}
{\large\bf VI. Application to the nonleptonic decays $\Omega_{b} \rightarrow
\Omega_{c}^{(*)} \;P \;(V)$}
\vspace{0.2in}

In this section we will apply the numerical solutions of the 
B-S equation to the nonleptonic decays $\Omega_{b} \rightarrow 
\Omega_{c}^{(*)}$ and a pseudoscalar or vector meson. In fact, 
$\Sigma_{b}^{(*)}$ and $\Xi_{b}^{(*)}$ decay strongly and their weak decays
are hard to observe. However, $\Omega_{b}$ decays only weakly. 
We will first calculate the Isgur-Wise functions $\xi(\omega)$ and 
$\zeta(\omega)$ for $\Omega_{b}^{(*)} \rightarrow\Omega_{c}^{(*)}$
in Eq.(\ref{4a}) and then apply them to the nonleptonic
weak decays of $\Omega_{b}$.

\vspace{0.2in}
{\large\bf A.  Isgur-Wise functions for $\Omega_{b}^{(*)} 
\rightarrow\Omega_{c}^{(*)}$}
\vspace{0.2in}

The Isgur-Wise functions $\xi(\omega)$ and $\zeta(\omega)$ are related to the
overlap integrals of the B-S wave functions of 
the initial ($\Omega_b$) and final
($\Omega_{c}^{(*)}$) states. The concrete expression for them can be 
obtained by comparing the structure  
$\bar{B}_{m'}^{\mu}(v')\Gamma B_{m \mu}(v)$ and
$v \cdot \bar{B}_{m'}(v')\Gamma v' \cdot B_{m}(v)$ on both sides of 
Eq.(\ref{4a}). Similarly to Eq.(\ref{4d}), we have the following equation after
substituting Eq.(\ref{2g})
into Eq.(\ref{4bb}) and using Eq.(\ref{4a})
\begin{eqnarray}
\bar{B}_{m'}^{\nu}(v')\Gamma
B_{m}^{\mu}(v)(\xi(\omega)g_{\mu\nu}+\zeta(\omega)v_\nu v'_\mu)&=&\int
\frac{\rd^4p}{(2\pi)^4}\frac{i}
{p_l+E_0+m_D+i\epsilon} \bar{\chi}_{Pm'}^{\mu}(p')\Gamma \nn\\
&&\int\frac{\rd^4q}{(2\pi)^4}G_{\mu\nu}(P,p,q)
\chi_{Pm}^{\nu}(p).
\label{6a}
\vspace{2mm}
\end{eqnarray}
Substituting Eq.(\ref{2o}) and Eq.(\ref{2p})
into Eq.(\ref{6a}) and using  Eqs.(\ref{3d}), (\ref{3e})
and (\ref{3ee}), we find that on the right hand side of 
Eq.(\ref{6a}) there are the following structures: 
$p'_t \cdot \bar{B}_{m'}(v')\Gamma v' \cdot B_{m}(v)$, 
$v \cdot \bar{B}_{m'}(v')\Gamma p_t \cdot B_{m}(v)$,
$p'_t \cdot \bar{B}_{m'}(v')\Gamma p_t \cdot B_{m}(v)$,
$p_t \cdot \bar{B}_{m'}(v')\Gamma p_t \cdot B_{m}(v)$, and
$p'_t \cdot \bar{B}_{m'}(v')\Gamma p'_t \cdot B_{m}(v)$. However, all of 
them can be expressed in terms of $\bar{B}_{m'}^{\mu}(v')\Gamma B_{m \mu}(v)$ 
and $v \cdot \bar{B}_{m'}(v')\Gamma v' \cdot B_{m}(v)$, after the integration 
over $p$, on the grounds of Lorentz invariance. Take 
$p'_t \cdot \bar{B}_{m'}(v')\Gamma p_t \cdot B_{m\mu}(v)$ as an example. In 
general, the integral $\int \frac{\rd^4 q}{(2\pi)^4}p_{t}^{\prime\mu}
p_{t}^{\nu}f$, where $f$ is some Lorentz scalar function, can be expressed in
terms of $g^{\mu\nu}, \; v^\mu v^\nu,\; v^{\prime\mu} v^{\prime\nu}, \;
v^{\prime\mu} v^{\nu}$ and $v^{\mu} v^{\prime\nu}$. However, only the
$g^{\mu\nu}$ and $v^{\mu} v^{\prime\nu}$ terms contribute when contracted
with $\bar{B}_{m'\mu}(v')\Gamma B_{m \nu}(v)$, leading to the
structures $\bar{B}_{m'}^{\mu}(v')\Gamma B_{m \mu}(v)$ and
$v \cdot \bar{B}_{m'}(v')\Gamma v' \cdot B_{m}(v)$, respectively. The
coefficients of these two terms can be obtained directly. In this way,
we have the following replacement rule:
\begin{eqnarray}
p'_t \cdot \bar{B}_{m'}(v')&\Gamma& v' \cdot B_{m}(v) \rightarrow
\frac{1}{1-\omega^2} v \cdot p'_t v \cdot \bar{B}_{m'}(v')\Gamma v' 
\cdot B_{m}(v), \nn\\
v \cdot \bar{B}_{m'}(v')&\Gamma& p_t \cdot B_{m}(v) \rightarrow
\frac{1}{1-\omega^2} v' \cdot p_t v \cdot \bar{B}_{m'}(v')\Gamma v' 
\cdot B_{m}(v), \nn\\
p'_t \cdot \bar{B}_{m'}(v')&\Gamma& p_t \cdot B_{m}(v) \rightarrow
\left[\frac{1}{2} p_t \cdot p'_t -\frac{\omega}{2(\omega^2-1)} v\cdot p'_t
v' \cdot p_t \right]\bar{B}_{m'}^{\mu}(v')\Gamma B_{m \mu}(v) \nn\\
&+&\left[-\frac{\omega}{2(\omega^2-1)} p_t \cdot p'_t +\frac{\omega^2+2}
{2(\omega^2-1)^2} v\cdot p'_t v' \cdot p_t \right]v \cdot \bar{B}_{m'}(v')
\Gamma v' \cdot B_{m}(v), \nn\\
p_t \cdot \bar{B}_{m'}(v')&\Gamma& p_t \cdot B_{m}(v) \rightarrow
\left[-\frac{1}{2} p_{t}^{2} +\frac{1}{2(\omega^2-1)} 
(v' \cdot p_t)^2 \right]\bar{B}_{m'}^{\mu}(v')\Gamma B_{m \mu}(v) \nn\\
&+&\left[\frac{\omega}{2(\omega^2-1)} p_{t}^{2} -\frac{3\omega}
{2(\omega^2-1)^2} (v' \cdot p_t)^2 \right]v \cdot \bar{B}_{m'}(v')
\Gamma v' \cdot B_{m}(v), \nn\\
p'_t \cdot \bar{B}_{m'}(v')&\Gamma& p'_t \cdot B_{m}(v) \rightarrow
\left[-\frac{1}{2} p_{t}^{\prime 2} +\frac{1}{2(\omega^2-1)} 
(v \cdot p'_t)^2 \right]\bar{B}_{m'}^{\mu}(v')\Gamma B_{m \mu}(v) \nn\\
&+&\left[\frac{\omega}{2(\omega^2-1)} p_{t}^{\prime 2} -\frac{3\omega}
{2(\omega^2-1)^2} (v \cdot p'_t)^2 \right]v \cdot \bar{B}_{m'}(v')
\Gamma v' \cdot B_{m}(v).
\label{6b}
\vspace{2mm}
\end{eqnarray}

Since in the weak transition the diquark acts as a spectator its
momentum in the initial and final baryons should
be the same, $p_2=p'_2$. Then we can show that
\begin{equation}
p'=p+m_D(v'-v),
\label{6c}
\vspace{2mm}
\end{equation}
where we have omitted the $O(1/m_Q)$ corrections. From Eq. (\ref{6c}) 
we have the following relations straightforwardly
\begin{eqnarray}
p'_l &=& p_l\omega -p_t \sqrt{\omega^2-1}{\rm cos}\theta, \nn \\
p^{'2}_{t} &=& p_{t}^{2}+p_{t}^{2}(\omega^2-1){\rm cos}^2\theta
+p_{l}^2(\omega^2-1) -2p_l p_t \omega \sqrt{\omega^2-1} {\rm cos}\theta,\nn\\
v' \cdot p_t &=& -p_t \sqrt{\omega^2 -1} {\rm cos}\theta, \nn\\
v \cdot p'_t &=& (1-\omega^2)p_l +\omega p_t \sqrt{\omega^2 -1}
{\rm cos}\theta, \nn\\
p_t \cdot p'_t &=& -p_{t}^{2}-p_{t}^{2}(\omega^2-1){\rm cos}^2\theta+\omega
\sqrt{\omega^2-1}p_l p_t {\rm cos}\theta,
\label{6d}
\vspace{2mm}
\end{eqnarray}
where $\theta$ is the angle between $p_t$ and $v'_t$.

With the aid of the relations between $A(p), \; C(p),\; D(p)$ and
$\tilde{A}(p_{t}^{2}), \; \tilde{C}(p_{t}^{2}),\; \tilde{D}(p_{t}^{2})$
[Eqs.(\ref{3k})-(\ref{3m})] and using Eqs.(\ref{6b}), (\ref{6d}) and the
integration formulae in Appendix A we have
the explicit expressions for $\xi (\omega)$ and  $\zeta (\omega)$ after 
integrating the $p_l$ component by selecting the proper contour,
\begin{eqnarray}
\xi (\omega) &=& \int \frac{p_{t}^{2}\rd p_t}{4\pi^2}\int_{0}^{\pi}
{\rm sin}\theta \rd \theta \frac{-1}{2W_p (E_0+m_D-W_p)(E_0+m_D-\omega W_p -p_t
\sqrt{\omega^2-1}{\rm cos}\theta)} \nn\\
&&\{-2W_p (E_0+m_D-W_p)F_A(p_{t}^{2}, {\rm cos}\theta) 
\tilde{A}(p_{t}^{2})-\frac{3}{4}
(1-{\rm cos}^2\theta)[h_1(|p_{t}|)-h_3(|p_{t}|)] \nn\\
&&F_A(p_{t}^{2}, {\rm cos}\theta)-\frac{1}{2}\omega p_{t}^{2}
(1-{\rm cos}^2\theta)
F_C(p_{t}^{2}, {\rm cos}\theta)
h_2(|p_{t}|)+\frac{3}{4}p_t \sqrt{\omega^2-1}(1-{\rm cos}^2\theta)\nn\\
&&{\rm cos}\theta [h_1(|p_{t}|)-h_3(|p_{t}|)]F_C(p_{t}^{2}, 
{\rm cos}\theta)+W_p
(1-{\rm cos}^2\theta)(E_0+m_D-W_p)p_{t}^{2}\tilde{A}(p_{t}^{2})\nn\\
&&F_D(p_{t}^{2}, {\rm cos}\theta)-\frac{1}{2}p_{t}^{2}
(1-{\rm cos}^2\theta)[(\omega^2-1)
W_p+\omega p_t \sqrt{\omega^2-1}{\rm cos}\theta]
h_2(|p_{t}|)\nn\\
&&F_D(p_{t}^{2}, {\rm cos}\theta)
+\frac{3}{4}(1-{\rm cos}^2\theta)[p_{t}^{2}+p_{t}^{2}(\omega^2-1)
{\rm cos}^2\theta
+\omega\sqrt{\omega^2-1}W_p p_t] \nn\\
&&F_D(p_{t}^{2}, {\rm cos}\theta)[h_1(|p_{t}|)-h_3(|p_{t}|)]\},
\label{6e}
\vspace{2mm}
\end{eqnarray}
and
\begin{eqnarray}
\zeta (\omega) &=& \int \frac{p_{t}^{2}\rd p_t}{4\pi^2}\int_{0}^{\pi}
{\rm sin}\theta \rd \theta \frac{-1}{2W_p (E_0+m_D-W_p)(E_0+m_D-\omega W_p -p_t
\sqrt{\omega^2-1}{\rm cos}\theta)} \nn\\
&&\left\{\frac{1}{\sqrt{\omega^2-1}}p_t {\rm cos}\theta 
F_A(p_{t}^{2}, {\rm cos}\theta)
h_2(|p_{t}|)+\frac{3\omega}{4(\omega^2-1)}(1-3{\rm cos}^2\theta)\right.\nn\\
&&\left. [h_1(|p_{t}|)-h_3(|p_{t}|)]F_A(p_{t}^{2}, {\rm cos}\theta)
+\left[W_p+\frac{\omega}{\sqrt{\omega^2-1}}p_t {\rm cos}\theta\right]
F_C(p_{t}^{2}, {\rm cos}\theta)\right.\nn\\
&&\left. 2W_p (E_0+m_D-W_p)\tilde{A}(p_{t}^{2})+\frac{\omega}{2(\omega^2-1)}
[\omega p_{t}^{2}-3\omega {\rm cos}^2\theta p_{t}^{2}\right.\nn\\
&&\left. -2W_p p_t \sqrt{\omega^2-1}{\rm cos}\theta]
F_C(p_{t}^{2}, {\rm cos}\theta)
h_2(|p_{t}|)-\frac{3}{4\sqrt{\omega^2-1}}[\omega p_{t}-3\omega 
{\rm cos}^2\theta 
p_{t} \right.\nn\\
&&\left.-2W_p\sqrt{\omega^2-1}{\rm cos}\theta]{\rm cos}\theta
[h_1(|p_{t}|)-h_3(|p_{t}|)]F_C(p_{t}^{2}, {\rm cos}\theta)
+\left[-\omega W_{p}^{2}\right.\right.\nn\\
&&\left.\left.-\frac{2\omega^2}{\sqrt{\omega^2-1}}W_p p_t{\rm cos}\theta
-\frac{\omega(2\omega^2+1)}{2(\omega^2-1)}p_{t}^{2}{\rm cos}^2\theta
+\frac{\omega}{2(\omega^2-1)}p_{t}^{2}\right]\right.\nn\\
&&\left.[-2W_p (E_0+m_D-W_p)]F_D(p_{t}^{2}, {\rm cos}\theta)
\tilde{A}(p_{t}^{2})
+\frac{1}{2\sqrt{\omega^2-1}}[\omega p_{t}^{2}-3\omega 
{\rm cos}^2\theta p_{t}^{2}
\right.\nn\\
&&\left.-2W_p p_t \sqrt{\omega^2-1}{\rm cos}\theta](\sqrt{\omega^2-1}W_p
+\omega p_t {\rm cos}\theta)F_D(p_{t}^{2}, {\rm cos}\theta)
h_2(|p_{t}|)\right.\nn\\
&&\left.-\frac{3}{4(\omega^2-1)}[\omega p_{t}-3\omega {\rm cos}^2\theta 
p_{t}-2W_p\sqrt{\omega^2-1}{\rm cos}\theta][p_t+p_t (\omega^2-1)
{\rm cos}^2\theta
\right.\nn\\
&&\left.+\omega\sqrt{\omega^2-1}W_p {\rm cos}\theta]
F_D(p_{t}^{2}, {\rm cos}\theta)
[h_1(|p_{t}|)-h_3(|p_{t}|)]\right\},
\label{6f}
\vspace{2mm}
\end{eqnarray}
where $h_i(|p_t|) (i=1,2,3,4)$ are given in Eqs.(\ref{4f})-(\ref{4i})
and $F_A(p_{t}^{2}, {\rm cos}\theta),\;F_C(p_{t}^{2}, {\rm cos}\theta)$ and
$F_D(p_{t}^{2}, {\rm cos}\theta)$ have the following expressions
\begin{eqnarray}
F_A(p_{t}^{2}, {\rm cos}\theta) &=& \int \frac{q_{t}^{2}\rd q_t}{4\pi^2}
\left\{8\pi\kappa F_1(|p'_{t}|,|q_{t}|)[\tilde{A}(q_{t}^{2})-\frac{1}{2}
q_{t}^{2}\tilde{D}(q_{t}^{2})-\tilde{A}(p_{t}^{\prime 2})]
+\frac{16\pi\beta}{3(Q_{1}^{2}-\mu^2)}\right. \nn\\
&&\left. [F_2(|p'_{t}|,|q_{t}|,\mu)-F_2(|p'_{t}|,|q_{t}|,Q_1)]
[2(\omega W_p+p_t\sqrt{\omega^2-1}{\rm cos}\theta)\right.\nn\\
&&\left.(\tilde{A}(q_{t}^{2})-\frac{1}{2}
q_{t}^{2}\tilde{D}(q_{t}^{2}))-\frac{1}{2}
q_{t}^{2}\tilde{C}(q_{t}^{2})]+\frac{1}{2p_{t}^{\prime 2}}
\frac{16\pi\beta}{3(Q_{1}^{2}-\mu^2)}[-F_4(|p'_{t}|,|q_{t}|,\mu)\right.\nn\\
&&\left.+F_4(|p'_{t}|,|q_{t}|,Q_1)][\tilde{C}(q_{t}^{2})+
2(\omega W_p+p_t\sqrt{\omega^2-1}{\rm cos}\theta)
\tilde{D}(q_{t}^{2})]\right.\nn\\
&&\left.+\frac{1}{2p_{t}^{\prime 2}}8\pi\kappa F_5(|p'_{t}|,|q_{t}|)
\tilde{D}(q_{t}^{2})\right\},
\label{6g}
\vspace{2mm}
\end{eqnarray}
\begin{eqnarray}
F_C(p_{t}^{2}, {\rm cos}\theta) &=& \frac{1}{m_{D}^{2}}
\int \frac{q_{t}^{2}\rd q_t}{4\pi^2}
\left\{8\pi\kappa F_1(|p'_{t}|,|q_{t}|)[(\omega W_p +p_t \sqrt{\omega^2-1}
{\rm cos}\theta)(\tilde{A}(q_{t}^{2})-\tilde{A}(p_{t}^{\prime 2})
\right.\nn\\
&&\left.+p_{t}^{\prime 2}\tilde{D}(p_{t}^{\prime 2}))+((\omega W_p +p_t 
\sqrt{\omega^2-1}{\rm cos}\theta)^2-m_{D}^{2})\tilde{C}(p_{t}^{\prime 2})]
+\frac{16\pi\beta}{3(Q_{1}^{2}-\mu^2)}\right.\nn\\
&&\left. [F_2(|p'_{t}|,|q_{t}|,\mu)-F_2(|p'_{t}|,|q_{t}|,Q_1)]
[(\omega W_p +p_t \sqrt{\omega^2-1}{\rm cos}\theta)^2+m_{D}^{2}]\right.\nn\\
&&\left.\tilde{A}(q_{t}^{2})-\frac{1}{p_{t}^{\prime 2}}
8\pi\kappa F_3(|p'_{t}|,|q_{t}|)[(\omega W_p +p_t 
\sqrt{\omega^2-1}{\rm cos}\theta)^2-m_{D}^{2}]\tilde{C}(q_{t}^{2})\right.\nn\\
&&\left.-\frac{1}{p_{t}^{\prime 2}}\frac{16\pi\beta}{3(Q_{1}^{2}-\mu^2)}
[-F_4(|p'_{t}|,|q_{t}|,\mu)+F_4(|p'_{t}|,|q_{t}|,Q_1)][(\omega W_p \right.\nn\\
&&\left.+p_t 
\sqrt{\omega^2-1}{\rm cos}\theta)\tilde{C}(q_{t}^{2})+
((\omega W_p +p_t \sqrt{\omega^2-1}{\rm cos}\theta)^2+m_{D}^{2})
\tilde{D}(q_{t}^{2})]
\right.\nn\\
&&\left.-\frac{1}{p_{t}^{\prime 2}}8\pi\kappa F_5(|p'_{t}|,|q_{t}|)
(\omega W_p +p_t \sqrt{\omega^2-1}{\rm cos}\theta)\tilde{D}(q_{t}^{2})\right\},
\label{6h}
\vspace{2mm}
\end{eqnarray}
\begin{eqnarray}
F_D(p_{t}^{2}, {\rm cos}\theta) &=& -\frac{1}{m_{D}^{2}}
\int \frac{q_{t}^{2}\rd q_t}{4\pi^2}
\left\{8\pi\kappa F_1(|p'_{t}|,|q_{t}|)\left[\tilde{A}(q_{t}^{2})+
\frac{m_{D}^{2}}{2p_{t}^{\prime 2}}q_{t}^{2}\tilde{D}(q_{t}^{2})
-\tilde{A}(p_{t}^{\prime 2})\right.\right. \nn\\
&&\left.\left.+(\omega W_p+p_t\sqrt{\omega^2-1}{\rm cos}\theta)
\tilde{C}(p_{t}^{\prime 2})+(p_{t}^{\prime 2}+m_{D}^{2})
\tilde{D}(p_{t}^{\prime 2})\right]
+\frac{16\pi\beta}{3(Q_{1}^{2}-\mu^2)}\right.\nn\\
&&\left.[F_2(|p'_{t}|,|q_{t}|,\mu)-F_2(|p'_{t}|,|q_{t}|,Q_1)]
\left[(\omega W_p +p_t \sqrt{\omega^2-1}{\rm cos}\theta)
\left(\tilde{A}(q_{t}^{2})
\right.\right.\right.\nn\\
&&\left.\left.\left.+\frac{m_{D}^{2}}{p_{t}^{\prime 2}}q_{t}^{2}
\tilde{D}(q_{t}^{2})\right)
+ \frac{m_{D}^{2}}{2p_{t}^{\prime 2}}q_{t}^{2}\tilde{C}(q_{t}^{2})\right]
-\frac{1}{p_{t}^{\prime 2}}8\pi\kappa F_3(|p'_{t}|,|q_{t}|)
(\omega W_p \right. \nn\\
&&\left. +p_t \sqrt{\omega^2-1}{\rm cos}\theta)\tilde{C}(q_{t}^{2})
-\frac{1}{p_{t}^{\prime 2}}\frac{16\pi\beta}{3(Q_{1}^{2}-\mu^2)}
[-F_4(|p'_{t}|,|q_{t}|,\mu)\right. \nn\\
&&\left. +F_4(|p'_{t}|,|q_{t}|,Q_1)]\left[\frac{3m_{D}^{2}+2p_{t}^{\prime 2}}
{2p_{t}^{\prime 2}}(\tilde{C}(q_{t}^{2})
+2(\omega W_p+p_t\sqrt{\omega^2-1}{\rm cos}\theta)\right.\right.\nn\\
&&\left.\left.\tilde{D}(q_{t}^{2}))
-(\omega W_p+p_t\sqrt{\omega^2-1}{\rm cos}\theta)\tilde{D}(q_{t}^{2})\right]
-\frac{3m_{D}^{2}+2p_{t}^{\prime 2}}{2p_{t}^{\prime 2}}\right.\nn\\
&&\left.8\pi\kappa F_5(|p'_{t}|,|q_{t}|)[\tilde{D}(q_{t}^{2})\right\}.
\label{6i}
\vspace{2mm}
\end{eqnarray}

	In Section V we obtained the numerical results for 
$\tilde{A}(p_{t}^{2})$, $\tilde{C}(p_{t}^{2})$, and $\tilde{D}(p_{t}^{2})$.
Substituting these results into Eqs.(\ref{6e}) and (\ref{6f}) we have the
numerical solutions for $\xi (\omega)$ and $\zeta (\omega)$ depending on
the parameters in our model. 
For $\Omega_{b}^{(*)} \rightarrow \Omega_{c}^{(*)}$, we show the Isgur-Wise 
functions with typical value $m_D=1.20$GeV in Fig.5(a) ($\kappa=0.02$GeV$^3$)
and Fig.5(b) ($\kappa=0.10$GeV$^3$) respectively. The dependence of the 
Isgur-Wise functions on $m_D$ are shown in 
Fig.5(c) ($m_D=1.15$GeV) and Fig.5(d) ($m_D=1.25$GeV) for 
$\kappa=0.06$GeV$^3$. It can be seen from these plots that 
$\xi (\omega)$ and $\zeta (\omega)$ have opposite signs and 
$\xi (\omega)$ changes more rapidly than $\zeta (\omega)$ as $\omega$
increases. 

	It is interesting to study the relation between $\xi (\omega)$ 
and $\zeta (\omega)$. Based on the picture that in the large $N_c$ limit heavy
baryons are viewed as the bound states of chiral solitons and heavy 
mesons\cite{jenkins}, Chow has shown that $\xi (\omega)$ and $\zeta (\omega)$
obey the following relation\cite{chow}
\begin{equation}
\xi (\omega)=-(1+\omega)\zeta (\omega).
\label{6j}
\vspace{2mm}
\end{equation}
The deviation from this relation is caused by $1/N_c$ corrections.
From Figs.5(a)-(d) we can see that, in the range of the parameters in our
model, Eq.(\ref{6j}) is generally satisfied. For some sets of parameters this
relation holds well.

\vspace{0.2in}
{\large\bf B. Nonleptonic decays $\Omega_{b} \rightarrow
\Omega_{c}^{(*)} \;P \;(V)$}
\vspace{0.2in}

In this section we will discuss the Cabbibo-allowed two body nonleptonic
decays of $\Omega_{b}\rightarrow \Omega_{c}^{(*)} \;P \;(V)$ ($P$ and $V$
stand for pseudoscalar and vector mesons respectively). 
The Hamiltonian describing such decays reads
\begin{equation}
H_{\rm eff}=\frac{G_F}{\sqrt{2}}V_{cb}V^{*}_{UD}(a_1 O_1 +a_2 O_2),
\label{6k}
\vspace{2mm}
\end{equation}
with $O_1=(\bar D U)(\bar c b)$ and $O_2=(\bar c U)(\bar D b)$, where $U$
and $D$ are the fields for light quarks involved in the decay, and 
$(\bar q_1 q_2)=\bar q_1 \gamma_\mu (1-\gamma_5) q_2$ is understood.
The parameters $a_1$ and $a_2$ are treated as free parameters since they
involve hadronization effects. Since $\Omega_{b}$ decays are energetic, 
the factorization assumption is applied so that one of the
currents in the Hamiltonian (\ref{6k}) is factorized out and
generates a meson\cite{bjorken, dugan}. 
Thus the decay amplitude of the two body nonleptonic decay
becomes the product of two matrix elements, one is related to the decay
constant of the factorized meson ($P$ or $V$) and the other is the weak 
transition matrix element between $\Omega_{b}$ and $\Omega_{c}^{(*)}$,
\begin{equation}
M^{fac}(\Omega_{b} \rightarrow \Omega_{c}^{(*)} P(V))=\frac{G_F}{\sqrt{2}}
V_{cb}V^{*}_{UD}
a_1 \langle P(V)|A_\mu(V_\mu)|0\rangle \langle\Omega_{c}^{(*)} (v')
|J^\mu|\Omega_{b}(v)\rangle.
\label{6l}
\vspace{2mm}
\end{equation}
Here $J_\mu$ is the weak current $(\bar c b)$ and $\langle 0|A_\mu(V_\mu)|P(V)
\rangle$ are related
to the decay constants of the pseudoscalar or vector mesons  by
\begin{eqnarray}
\langle 0|A_\mu|P\rangle&=&if_P q_\mu, \nn\\
\langle 0|V_\mu|V\rangle&=&f_V m_V \epsilon_{\mu},
\label{6m}
\vspace{2mm}
\end{eqnarray}
where $q_\mu$ is the momentum of the emitted meson (from the W-boson),
$\epsilon_\mu$ is the polarization vector of the vector meson, and the
normalization for the decay constants is chosen so that $f_\pi=132$MeV. 
It is noted that in the two body nonleptonic weak decays
$\Omega_{b} \rightarrow \Omega_{c}^{(*)} P(V)$ 
there is no contribution from the $a_2$ term, since such a term corresponds 
to the transition of $\Omega_{b}$ to a light baryon instead of 
$\Omega_{c}^{(*)}$.
On the other hand, the general form for the amplitudes of 
$\Omega_{b} \rightarrow \Omega_{c}^{(*)} P(V)$ are 
\begin{eqnarray}
M(\Omega_{b} \rightarrow \Omega_{c} P)&=&i\bar{u}_{f}(v')
(A+B\gamma_5)u_{i}(v), \nn \\
M(\Omega_{b} \rightarrow \Omega_{c} V)&=&\bar{u}_{f}(v')
\epsilon^{*\mu}
[A_1\gamma_\mu \gamma_5+A_2 p_{f\mu}\gamma_5+B_1\gamma_\mu
+B_2 p_{f\mu}]u_{i}(v), \nn\\
M(\Omega_{b} \rightarrow \Omega_{c}^{*} P)&=&iq_\mu\bar{u}_{f}^{\mu}(v')
(C+D\gamma_5)u_{i}(v), \nn \\
M(\Omega_{b} \rightarrow \Omega_{c}^{*} V)&=&\bar{u}_{f}^{\nu}(v')
\epsilon^{*\mu}[g_{\nu\mu}(C_1+D_1\gamma_5)+p_{i\nu}\gamma_\mu(C_2+D_2\gamma_5)
\nn\\
&&+p_{i\nu}p_{f\mu}(C_3+D_3\gamma_5)]u_{i}(v), 
\label{6n}
\vspace{2mm}
\end{eqnarray}
where $u_{i}$ is the dirac spinor of $\Omega_{b}$, $u_{f}^{(\mu)}$ is the
dirac (Rarita-Schwinger) spinor of $\Omega_{c}^{(*)}$, and
$p_{i(f)}$ is the momentum of $\Omega_{b}$ ($\Omega_{c}^{(*)}$).

From Eqs.(\ref{6l})-(\ref{6n}) and using Eq.(\ref{4a}) we find
\begin{eqnarray}
A&=&\frac{G_F}{\sqrt{2}}V_{cb}V^{*}_{UD}a_1 f_P \frac{1}{3}(m_i-m_f)[(\omega+2)
\xi(\omega)+(\omega^2-1)\zeta (\omega)], \nn\\
B&=&\frac{G_F}{\sqrt{2}}V_{cb}V^{*}_{UD}a_1 f_P \frac{1}{3}(m_i+m_f)[(3\omega
-2)\xi(\omega)+3(\omega^2-1)\zeta (\omega)], \nn\\
A_1&=&B_1=\frac{G_F}{\sqrt{2}}V_{cb}V^{*}_{UD}a_1 f_V m_V \frac{1}{3}[\omega
\xi(\omega)+(\omega^2-1)\zeta (\omega)], \nn\\
A_2&=&\frac{G_F}{\sqrt{2}}V_{cb}V^{*}_{UD}a_1 f_V m_V \frac{2}{3}
\left(\frac{1}{m_i}-\frac{1}{m_f}\right)[\xi(\omega)+(\omega+1)
\zeta (\omega)], \nn\\
B_2&=&-\frac{G_F}{\sqrt{2}}V_{cb}V^{*}_{UD}a_1 f_V m_V \frac{2}{3}
\left(\frac{1}{m_i}+\frac{1}{m_f}\right)[\xi(\omega)+(\omega-1)
\zeta (\omega)], \nn\\
C&=&-\frac{G_F}{\sqrt{2}}V_{cb}V^{*}_{UD}a_1 f_P \frac{1}{\sqrt{3}}
\left(1+\frac{m_f}{m_i}\right)[\xi(\omega)+(\omega-1)\zeta (\omega)], \nn\\
D&=&-\frac{G_F}{\sqrt{2}}V_{cb}V^{*}_{UD}a_1 f_P \frac{1}{\sqrt{3}}
\left[\left(1-\frac{m_f}{m_i}\right)\xi(\omega)+\left(\omega-1
-(\omega+3)\frac{m_f}{m_i}\right)\zeta (\omega)\right], \nn\\
C_1&=&D_1=\frac{G_F}{\sqrt{2}}V_{cb}V^{*}_{UD}a_1 f_V m_V \frac{2}{\sqrt{3}}
\xi(\omega),\nn\\
C_2&=&-\frac{G_F}{\sqrt{2}}V_{cb}V^{*}_{UD}a_1 f_V m_V \frac{1}{\sqrt{3}}
\frac{1}{m_i}[\xi(\omega)+(\omega+1)\zeta (\omega)],\nn\\
D_2&=&\frac{G_F}{\sqrt{2}}V_{cb}V^{*}_{UD}a_1 f_V m_V \frac{1}{\sqrt{3}}
\frac{1}{m_i}[\xi(\omega)+(\omega-1)\zeta (\omega)],\nn\\
C_3&=&D_3=\frac{G_F}{\sqrt{2}}V_{cb}V^{*}_{UD}a_1 f_V m_V \frac{2}{\sqrt{3}}
\frac{1}{m_i m_f}\zeta(\omega),
\label{6o}
\vspace{2mm}
\end{eqnarray}
where $m_i$ ($m_f$) is the mass of $\Omega_{b}$ ($\Omega_{c}^{(*)}$).

With Eqs.(\ref{6n}) and (\ref{6o}) we can calculate the decay widths and
polarization parameters for $\Omega_{b}\rightarrow \Omega_{c}^{(*)} \;P \;(V)$.
The kinematic formulae which have been derived using both 
partial wave and helicity
methods can be found in references\cite{tuan, korner}. These two methods are
equivalent. For instance, in the helicity method\cite{korner}, 
the decay width is expressed in terms of the helicity amplitudes,
\begin{equation}
\Gamma=\frac{p_c}{16\pi m_{i}^{2}}\sum_{\lambda_i,\lambda_f}
|h_{\lambda_f,\lambda_{P(V)};\lambda_i}|^2,
\label{6p}
\vspace{2mm}
\end{equation}
where $p_c$ is the c.m. momentum of the decay products and
the helicity amplitudes are defined as 
\begin{equation}
h_{\lambda_f,\lambda_{P(V)};\lambda_i}=\langle\Omega_{c}^{(*)}(\lambda_f),
P(V)(\lambda_{P(V)})| H_{\rm eff}|\Omega_{b}(\lambda_i)\rangle\;\;(\lambda_f-
\lambda_{P(V)}=\lambda_i).
\label{6q}
\vspace{2mm}
\end{equation}
The ``up-down'' asymmetry is given by
\begin{equation}
\alpha=\frac{\sum_{\lambda_f}(|h_{\lambda_f,\lambda_{P(V)};1/2}|^2-
|h_{\lambda_f,\lambda_{P(V)};-1/2}|^2)}{\sum_{\lambda_i,\lambda_f}
|h_{\lambda_f,\lambda_{P(V)};\lambda_i}|^2}.
\label{6r}
\vspace{2mm}
\end{equation}
The relations between the helicity amplitudes and the amplitudes given in 
Eq.(\ref{6n}), which we will not list here, can be found in 
\cite{korner, cheng}. Then from Eqs.(\ref{6o})-(\ref{6r}), we obtain the 
numerical results for the decay widths and asymmetry parameters. In Table 4
we list the results for $m_D=1.20$GeV. The numbers without (with) brackets
correspond to $\kappa=0.02$GeV$^3$ ($\kappa=0.10$GeV$^3$). 
The results for $\kappa=0.06$GeV$^3$ in the range $m_D=1.15$GeV (without
brackets) and  $m_D=1.25$GeV (with brackets) are shown in Table 5.
In the calculations we have taken $m_{\Omega_{b}}=6.14$GeV and the 
following decay constants $$f_\pi=132MeV, \;\;  f_{D_s}=241MeV\cite{sa},\;\;
f_\rho=216MeV, \;\;  f_{D_s}=f_{D_{s}^{*}}.$$

It can be seen from Tables 4 and 5 that the predictions for the decay widths
show a strong dependence on the parameters $\kappa$ and $m_D$ in our model. 
The experimental data in the future will be used to
fix these parameters and test our
model. However, the dependence of the up-down asymmetries on these 
parameters is slight. 

The decay widths and asymmetry parameters have also been calculated in the 
nonrelativistic quark
model approach\cite{cheng}, where the form factors are calculated at the
zero-recoil point and then extrapolated to other $\omega$ values under
the assumption of a dipole behaviour. 
By comparing the predictions in the B-S and
quark models we find that the decay widths in our model are smaller than
those in the quark model. For the asymmetry parameters, the difference is
even larger. Except for the processes 
$\Omega_{b}^{-}\rightarrow \Omega_{c}^{0 *} \pi^{-}$ and 
$\Omega_{b}^{-}\rightarrow \Omega_{c}^{0 *} D_{s}^{-}$ in Tables 4 and 5,
even the signs of $\alpha$ in these two models are opposite.

\begin{table}
\caption{Predictions for decay widths and asymmetry parameters for
$\Omega_{b}\rightarrow \Omega_{c}^{(*)} \;P \;(V)$ for $m_D=1.20$GeV.}
\begin{center}
\begin{tabular}{ccc}
\hline
\hline
Process &$\Gamma (10^{10}s^{-1})$ &$\alpha$ \\ 
\hline
$\Omega_{b}^{-}\rightarrow \Omega_{c}^{0} \pi^{-}$ &$0.052a_{1}^{2} 
\;\; (0.154a_{1}^{2})$ &$-0.67 \;\; (-0.70)$\\
\hline
$\Omega_{b}^{-}\rightarrow \Omega_{c}^{0} D_{s}^{-}$ &$0.261a_{1}^{2} 
 \;\;(0.592a_{1}^{2})$ &$-0.56  \;\;(-0.58)$\\
\hline
$\Omega_{b}^{-}\rightarrow \Omega_{c}^{0} \rho^{-}$ &$0.073a_{1}^{2} 
 \;\;(0.207a_{1}^{2})$ &$-0.68 \;\; (-0.71)$\\
\hline
$\Omega_{b}^{-}\rightarrow \Omega_{c}^{0} D_{s}^{* -}$ &$0.115a_{1}^{2} 
 \;\;(0.245a_{1}^{2})$ &$-0.73 \;\; (-0.74)$\\
\hline
$\Omega_{b}^{-}\rightarrow \Omega_{c}^{* 0} \pi^{-}$ &$0.046a_{1}^{2} 
 \;\;(0.133a_{1}^{2})$ &$-0.61  \;\;(-0.58)$\\
\hline
$\Omega_{b}^{-}\rightarrow \Omega_{c}^{* 0} D_{s}^{-}$ &$0.165a_{1}^{2} 
 \;\;(0.370a_{1}^{2})$ &$-0.54 \;\; (-0.52)$\\
\hline
$\Omega_{b}^{-}\rightarrow \Omega_{c}^{* 0} \rho^{-}$ &$0.134a_{1}^{2} 
 \;\;(0.354a_{1}^{2})$ &$0.59 \;\; (0.59)$\\
\hline
$\Omega_{b}^{-}\rightarrow \Omega_{c}^{* 0} D_{s}^{* -}$ &$0.462a_{1}^{2} 
 \;\;(0.960a_{1}^{2})$ &$0.31 \;\; (0.31)$\\
\hline
\hline
\end{tabular}
\end{center}
\end{table}

\begin{table}
\caption{Predictions for decay widths and asymmetry parameters for
$\Omega_{b}\rightarrow \Omega_{c}^{(*)} \;P \;(V)$ for $\kappa=0.06$GeV$^3$.}
\begin{center}
\begin{tabular}{ccc}
\hline
\hline
Process &$\Gamma (10^{10}s^{-1})$ &$\alpha$ \\ 
\hline
$\Omega_{b}^{-}\rightarrow \Omega_{c}^{0} \pi^{-}$ &$0.075a_{1}^{2} 
\;\; (0.145a_{1}^{2})$ &$-0.64 \;\; (-0.72)$\\
\hline
$\Omega_{b}^{-}\rightarrow \Omega_{c}^{0} D_{s}^{-}$ &$0.358a_{1}^{2} 
 \;\;(0.562a_{1}^{2})$ &$-0.54  \;\;(-0.59)$\\
\hline
$\Omega_{b}^{-}\rightarrow \Omega_{c}^{0} \rho^{-}$ &$0.102a_{1}^{2} 
 \;\;(0.150a_{1}^{2})$ &$-0.65 \;\; (-0.70)$\\
\hline
$\Omega_{b}^{-}\rightarrow \Omega_{c}^{0} D_{s}^{* -}$ &$0.149a_{1}^{2} 
 \;\;(0.232a_{1}^{2})$ &$-0.71 \;\; (-0.75)$\\
\hline
$\Omega_{b}^{-}\rightarrow \Omega_{c}^{* 0} \pi^{-}$ &$0.067a_{1}^{2} 
 \;\;(0.123a_{1}^{2})$ &$-0.64  \;\;(-0.56)$\\
\hline
$\Omega_{b}^{-}\rightarrow \Omega_{c}^{* 0} D_{s}^{-}$ &$0.227a_{1}^{2} 
 \;\;(0.345a_{1}^{2})$ &$-0.56 \;\; (-0.50)$\\
\hline
$\Omega_{b}^{-}\rightarrow \Omega_{c}^{* 0} \rho^{-}$ &$0.200a_{1}^{2} 
 \;\;(0.314a_{1}^{2})$ &$0.59 \;\; (0.59)$\\
\hline
$\Omega_{b}^{-}\rightarrow \Omega_{c}^{* 0} D_{s}^{* -}$ &$0.616a_{1}^{2} 
 \;\;(0.888a_{1}^{2})$ &$0.31 \;\; (0.31)$\\
\hline
\hline
\end{tabular}
\end{center}
\end{table}

\vspace{0.2in}
{\large\bf VII. Summary and discussion}
\vspace{0.2in}

Since in the heavy quark limit the light degrees of freedom in a heavy baryon
have good spin and isospin quantum numbers and since
the internal structure is blind to the
flavor and spin direction of the heavy quark, we assume that a heavy
baryon, $\omeq$, is composed of a heavy quark and a
light axial vector diquark. Based on this picture,
we establish the B-S equation for the heavy baryon $\omeq$. We discuss the
form of the B-S wave function and 
find that in the heavy quark limit there are three B-S scalar functions
to describe the dynamics inside a heavy baryon $\omeq$. This is consistent
with our physical picture. In order to solve the B-S equation, we 
assume a kernel containing a scalar confinement term and a one-gluon-exchange 
term, as in the $\Lambda_Q$ case. In the heavy quark limit, the heavy quark
is almost on mass-shell inside a heavy baryon and it is appropriate to
apply the covariant instantaneous approximation in the kernel.
Then we derive explicitly three coupled integral equations for the 
three B-S scalar functions $\tilde{A}$, $\tilde{C}$ and $\tilde{D}$.
These equations are solved numerically and we give the model predictions for
these functions. The results appear reasonable.
It is shown that the shapes of these functions are
similar for $\Sigma_{Q}^{(*)}$, $\Xi_{Q}^{(*)}$ and $\Omega_{Q}^{(*)}$,
with differences arising from $SU(3)$ flavor symmetry breaking effects. 

Although the B-S equation is formally the exact equation to describe
the bound state, there is much difficulty in applying it to the real
physical state. The most difficult point is that we must take a
phenomenologically inspired form for
%can not solve out the form of 
the kernel. 
%Hence we have to use some phenomenological kernel. 
Furthermore, we have used the quark and diquark propagators with
their free form, which
leads to some uncertainties. In our approach, there are
several parameters such as $\kappa$, $m_D$ and $\alpha_{s}^{({\rm eff})}$,
subject to the condition that the observed masses of $\omeq$ be reproduced. 
In our numerical solutions we let these parameters 
vary in some reasonable range. Another parameter is
$Q_{1}^{2}$, which arises from the internal structure of diquark. Its value for
the $(qq')$ diquark ($q,q' =u$ or $d$) is extracted from  
the data of the electromagnetic form factor of the proton.
When there is a strange quark in the diquark, we do not have a means to
determine its exact value 
at present. In the future, the experimental data for $\omeq$ should
help to fix the parameters in our model. 

As phenomenological applications, this model has been used to calculate the 
the Isgur-Wise functions $\xi(\omega)$ and $\zeta(\omega)$
for $\Omega_{b}^{(*)} \rightarrow \Omega_{c}^{(*)}$, 
and consequently, has provided theoretical
predictions for the Cabbibo-allowed
two body nonleptonic decay rates and up-down asymmetries
for the physical processes $\Omega_{b}\rightarrow \Omega_{c}^{(*)} \;P \;(V)$. 
It has been shown that the relation between
$\xi (\omega)$ and $\zeta (\omega)$ in our model is generally consistent
with that in the soliton model in the large $N_c$ limit.  
We have also compared our model predictions with those in the nonrelativistic
quark model. Our model yields decay widths which are much smaller and
for the asymmetry parameters the difference is even bigger. All these 
predictions will be tested in future experiments.

\vspace{1cm}

\noindent {\bf Acknowledgment}:

This work was supported in part by the Australian Research Council and
the National Science Foundation of China.

\vspace{2cm}

%\newpage

\vspace{0.2in}
{\large\bf Appendix A. Integration formulae}
\vspace{0.2in}

In this appendix we give the formulae which are used to reduce the
three dimensional
integration to the one dimensional integration. In the following formulae 
$\phi(q_{t}^{2})$ is some arbitrary function of $q_{t}^{2}$. The relevant
results needed are:

\begin{equation}
I_1=\int \frac{\rd^3 q_t}{(2\pi)^3}\frac{\phi(q_{t}^{2})}{[(p_t-q_t)^2
+\mu^2]^2}=\int
\frac{q_{t}^{2}\rd q_t}{4\pi^2}\phi(q_{t}^{2})F_1(|p_{t}|,|q_{t}|),
\label{aa}
\vspace{2mm}
\end{equation}
with
\begin{equation}
F_1(|p_{t}|,|q_{t}|)=\frac{2}{(p_{t}^{2}+q_{t}^{2}+\mu^2)^2
-4p_{t}^{2}q_{t}^{2}}.
\label{ab}
\vspace{2mm}
\end{equation}

\begin{equation}
I_2=\int \frac{\rd^3 q_t}{(2\pi)^3}\frac{\phi(q_{t}^{2})}{(p_t-q_t)^2
+\delta^2}=\int\frac{q_{t}^{2}\rd q_t}{4\pi^2}\phi(q_{t}^{2})F_2(|p_{t}|,
|q_{t}|,\delta),
\label{ac}
\vspace{2mm}
\end{equation}
with
\begin{equation}
F_2(|p_{t}|,|q_{t}|,\delta)=\frac{1}{2|p_{t}||q_{t}|}{\rm ln}
\frac{(|p_{t}|+|q_{t}|)^{2}+\delta^2}{(|p_{t}|-|q_{t}|)^{2}+\delta^2}.
\label{ad}
\vspace{2mm}
\end{equation}

\begin{equation}
I_3=\int \frac{\rd^3 q_t}{(2\pi)^3}\frac{p_t \cdot q_t
\phi(q_{t}^{2})}{[(p_t-q_t)^2
+\mu^2]^2}=\int
\frac{q_{t}^{2}\rd q_t}{4\pi^2}\phi(q_{t}^{2})F_3(|p_{t}|,|q_{t}|),
\label{ae}
\vspace{2mm}
\end{equation}
with
\begin{equation}
F_3(|p_{t}|,|q_{t}|)=\frac{1}{4|p_{t}||q_{t}|}\left[{\rm ln}
\frac{(|p_{t}|-|q_{t}|)^{2}+\mu^2}{(|p_{t}|+|q_{t}|)^{2}+\mu^2}
+\frac{4|p_{t}||q_{t}|(p_{t}^{2}+q_{t}^{2}+\mu^2)}{
(p_{t}^{2}+q_{t}^{2}+\mu^2)^2-4p_{t}^{2}q_{t}^{2}}\right].
\label{af}
\vspace{2mm}
\end{equation}

\begin{equation}
I_4=\int \frac{\rd^3 q_t}{(2\pi)^3}\frac{(p_t \cdot q_t)^2
\phi(q_{t}^{2})}{(p_t-q_t)^2
+\delta^2}=-\int\frac{q_{t}^{2}\rd q_t}{4\pi^2}\phi(q_{t}^{2})F_4(|p_{t}|,
|q_{t}|,\delta),
\label{ag}
\vspace{2mm}
\end{equation}
with
\begin{equation}
F_4(|p_{t}|,|q_{t}|,\delta)=\frac{p_{t}^{2}+q_{t}^{2}+\delta^2}
{2}\left[1+\frac{p_{t}^{2}+q_{t}^{2}+\delta^2}{4|p_{t}||q_{t}|}{\rm ln}
\frac{(|p_{t}|-|q_{t}|)^{2}+\delta^2}{(|p_{t}|+|q_{t}|)^{2}+\delta^2}\right].
\label{ah}
\vspace{2mm}
\end{equation}

\begin{equation}
I_5=\int \frac{\rd^3 q_t}{(2\pi)^3}\frac{(p_t \cdot q_t)^2
\phi(q_{t}^{2})}{[(p_t-q_t)^2
+\mu^2]^2}=\int
\frac{q_{t}^{2}\rd q_t}{4\pi^2}\phi(q_{t}^{2})F_5(|p_{t}|,|q_{t}|),
\label{ai}
\vspace{2mm}
\end{equation}
with
\begin{equation}
F_5(|p_{t}|,|q_{t}|)=\frac{1}{2}\left[1+\frac{p_{t}^{2}+q_{t}^{2}+\mu^2}
{2|p_{t}||q_{t}|}{\rm ln}
\frac{(|p_{t}|-|q_{t}|)^{2}+\mu^2}{(|p_{t}|+|q_{t}|)^{2}+\mu^2}
+\frac{(p_{t}^{2}+q_{t}^{2}+\mu^2)^2}{
(p_{t}^{2}+q_{t}^{2}+\mu^2)^2-4p_{t}^{2}q_{t}^{2}}\right].
\label{aj}
\vspace{2mm}
\end{equation}

\newpage

\vspace{0.2in}

{\large \bf Figure Captions} \\
\vspace{0.2in}

Fig.1 The vertex of two axial vector diquarks and a gluon.\\

\vspace{0.2cm}

Fig.2 The B-S scalar wave functions for $\Sigma_{Q}^{(*)}$. The units
are: GeV$^{-2}$ for  $\tilde{A}(p_{t}^{2})$, GeV$^{-3}$ for 
$\tilde{C}(p_{t}^{2})$ and GeV$^{-4}$ for $\tilde{D}(p_{t}^{2})$.
(a) For $m_D=0.95$GeV, we show the dependence on $|p_t|$ for two values of
$\kappa$. In the upper plane, the upper (lower) solid line is for 
$\tilde{A}(p_{t}^{2})$ ($\tilde{C}(p_{t}^{2})$)
with $\kappa=0.02$GeV$^3$, the upper (lower) dotted line is for 
$\tilde{A}(p_{t}^{2})$ ($\tilde{C}(p_{t}^{2})$) with $\kappa=0.10$GeV$^3$. 
In the lower plane, the solid line is for $\tilde{D}(p_{t}^{2})$ with 
$\kappa=0.02$GeV$^3$ and the dotted line is for $\tilde{D}(p_{t}^{2})$ with 
$\kappa=0.10$GeV$^3$.
(b) For $\kappa=0.06$GeV$^3$, we show
the dependence on $|p_t|$ for two values of $m_D$. 
In the upper plane, the upper (lower) solid line is for 
$\tilde{A}(p_{t}^{2})$ ($\tilde{C}(p_{t}^{2})$)
with $m_D=0.90$GeV, the upper (lower) dotted line is for 
$\tilde{A}(p_{t}^{2})$ ($\tilde{C}(p_{t}^{2})$) with $m_D=1.00$GeV.
In the lower plane, the solid line is for $\tilde{D}(p_{t}^{2})$ with 
$m_D=0.90$GeV and the dotted line is for $\tilde{D}(p_{t}^{2})$ with 
$m_D=1.00$GeV.\\

\vspace{0.2cm}

Fig.3 The B-S scalar wave functions for $\Xi_{Q}^{(*)}$. The units
are: GeV$^{-2}$ for  $\tilde{A}(p_{t}^{2})$, GeV$^{-3}$ for 
$\tilde{C}(p_{t}^{2})$ and GeV$^{-4}$ for $\tilde{D}(p_{t}^{2})$.
(a) For $m_D=1.15$GeV, we show the dependence on $|p_t|$ for two values of
$\kappa$. In the upper plane, the upper (lower) solid line is for 
$\tilde{A}(p_{t}^{2})$ ($\tilde{C}(p_{t}^{2})$)
with $\kappa=0.02$GeV$^3$, the upper (lower) dotted line is for 
$\tilde{A}(p_{t}^{2})$ ($\tilde{C}(p_{t}^{2})$) with $\kappa=0.10$GeV$^3$. 
In the lower plane, the solid line is for $\tilde{D}(p_{t}^{2})$ with 
$\kappa=0.02$GeV$^3$ and the dotted line is for $\tilde{D}(p_{t}^{2})$ with 
$\kappa=0.10$GeV$^3$.
(b) For $\kappa=0.06$GeV$^3$, we show
the dependence on $|p_t|$ for two values of $m_D$. 
In the upper plane, the upper (lower) solid line is for 
$\tilde{A}(p_{t}^{2})$ ($\tilde{C}(p_{t}^{2})$)
with $m_D=1.10$GeV, the upper (lower) dotted line is for 
$\tilde{A}(p_{t}^{2})$ ($\tilde{C}(p_{t}^{2})$) with $m_D=1.20$GeV.
In the lower plane, the solid line is for $\tilde{D}(p_{t}^{2})$ with 
$m_D=1.10$GeV and the dotted line is for $\tilde{D}(p_{t}^{2})$ with 
$m_D=1.20$GeV.\\

\vspace{0.2cm}

Fig.4 The B-S scalar wave functions for $\Omega_{Q}^{(*)}$. The units
are: GeV$^{-2}$ for  $\tilde{A}(p_{t}^{2})$, GeV$^{-3}$ for 
$\tilde{C}(p_{t}^{2})$ and GeV$^{-4}$ for $\tilde{D}(p_{t}^{2})$.
(a) For $m_D=1.20$GeV, we show the dependence on $|p_t|$ for two values of
$\kappa$. In the upper plane, the upper (lower) solid line is for 
$\tilde{A}(p_{t}^{2})$ ($\tilde{C}(p_{t}^{2})$)
with $\kappa=0.02$GeV$^3$, the upper (lower) dotted line is for 
$\tilde{A}(p_{t}^{2})$ ($\tilde{C}(p_{t}^{2})$) with $\kappa=0.10$GeV$^3$. 
In the lower plane, the solid line is for $\tilde{D}(p_{t}^{2})$ with 
$\kappa=0.02$GeV$^3$ and the dotted line is for $\tilde{D}(p_{t}^{2})$ with 
$\kappa=0.10$GeV$^3$.
(b) For $\kappa=0.06$GeV$^3$, we show
the dependence on $|p_t|$ for two values of $m_D$. 
In the upper plane, the upper (lower) solid line is for 
$\tilde{A}(p_{t}^{2})$ ($\tilde{C}(p_{t}^{2})$)
with $m_D=1.15$GeV, the upper (lower) dotted line is for 
$\tilde{A}(p_{t}^{2})$ ($\tilde{C}(p_{t}^{2})$) with $m_D=1.25$GeV.
In the lower plane, the solid line is for $\tilde{D}(p_{t}^{2})$ with 
$m_D=1.15$GeV and the dotted line is for $\tilde{D}(p_{t}^{2})$ with 
$m_D=1.25$GeV.\\
\vspace{0.2cm}

Fig.5(a)-(d) Numerical solutions for $\xi (\omega)$ and $\zeta (\omega)$ 
for $\Omega_{b}^{(*)} \rightarrow \Omega_{c}^{(*)}$. The upper solid line
is $\xi (\omega)$ and the lower solid line is $\zeta (\omega)$. The dotted
line is $-\xi (\omega)/(\omega+1)$. The parameters are $m_D=1.20$GeV and
$\kappa=0.02$GeV$^3$ in (a), $m_D=1.20$GeV and  $\kappa=0.10$GeV$^3$ in (b),
$\kappa=0.06$GeV$^3$ and  $m_D=1.15$GeV in (c), $\kappa=0.06$GeV$^3$ and  
$m_D=1.25$GeV in (d).

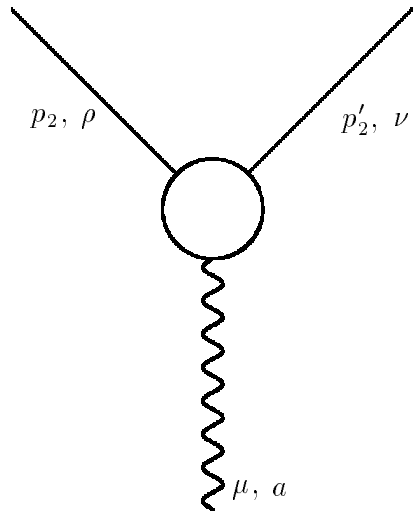
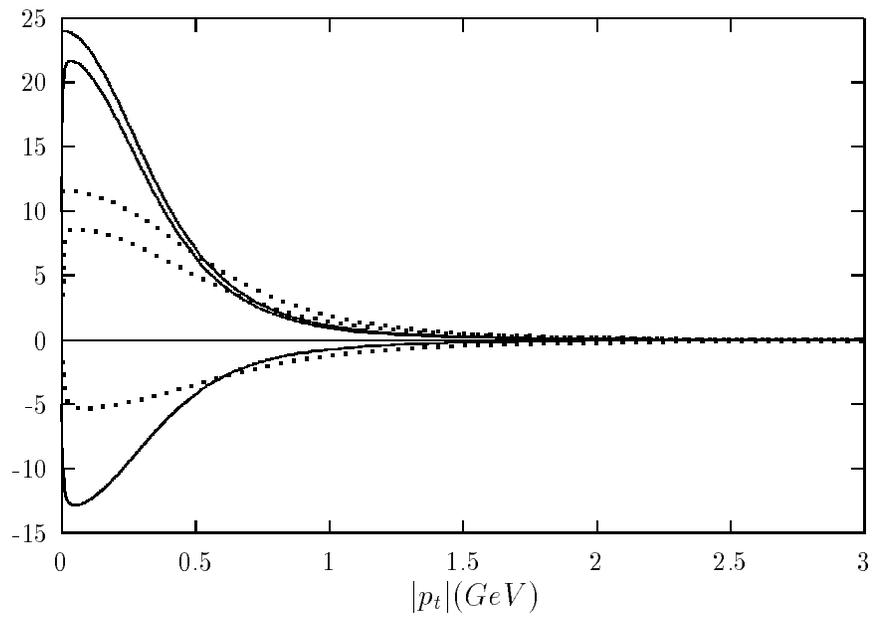
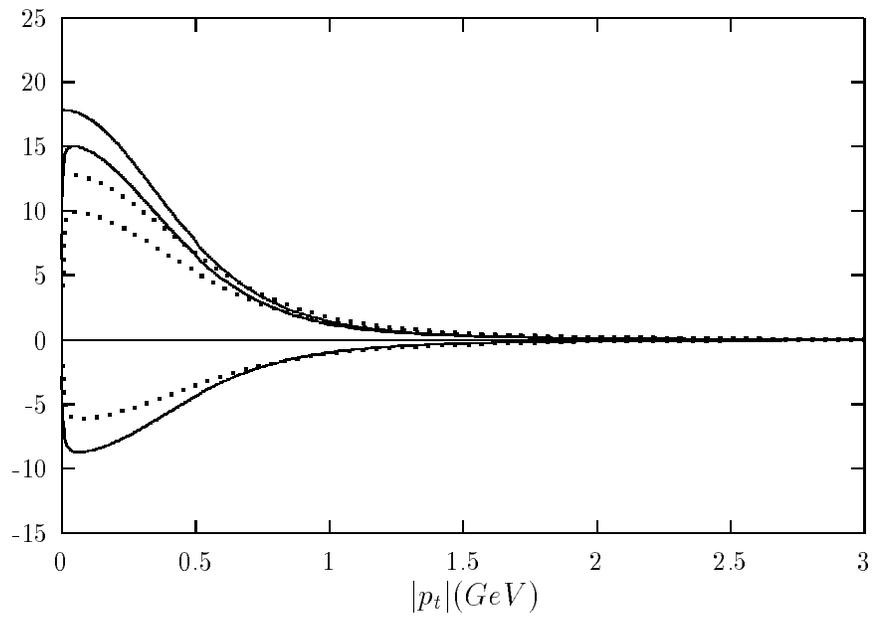
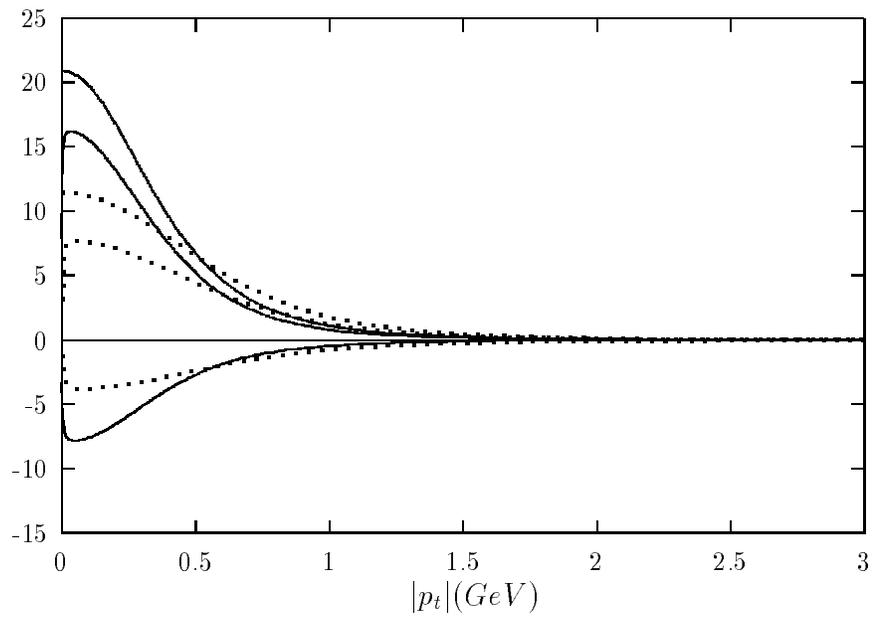
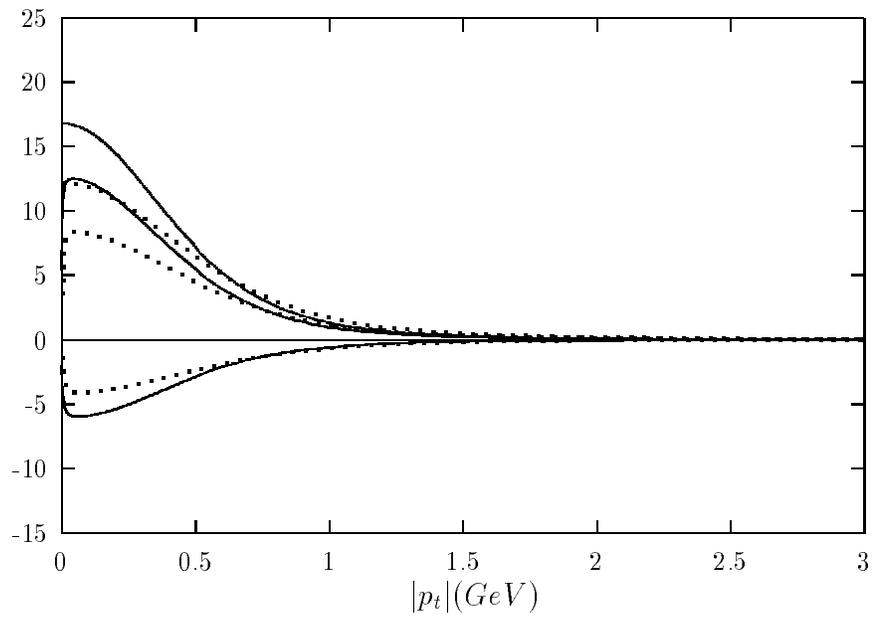
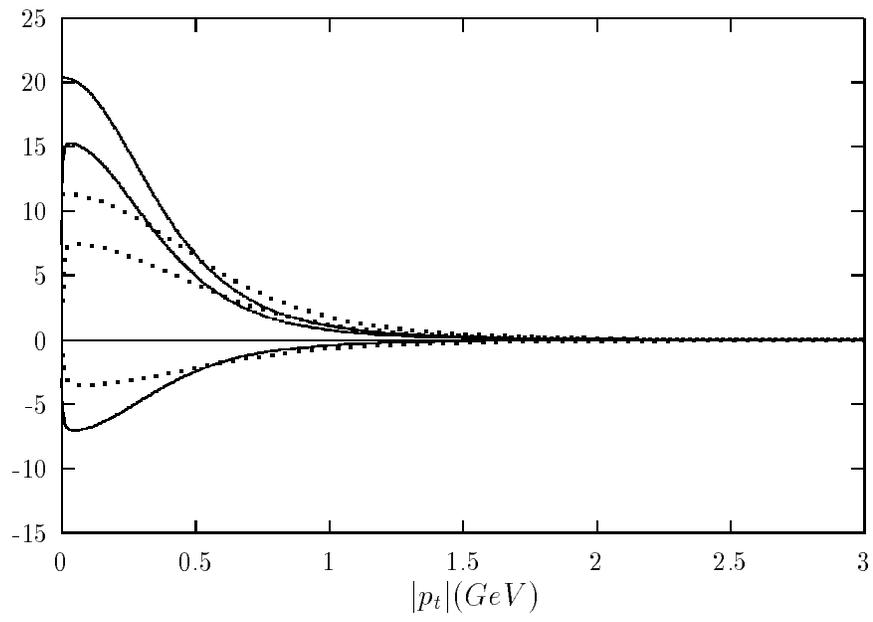
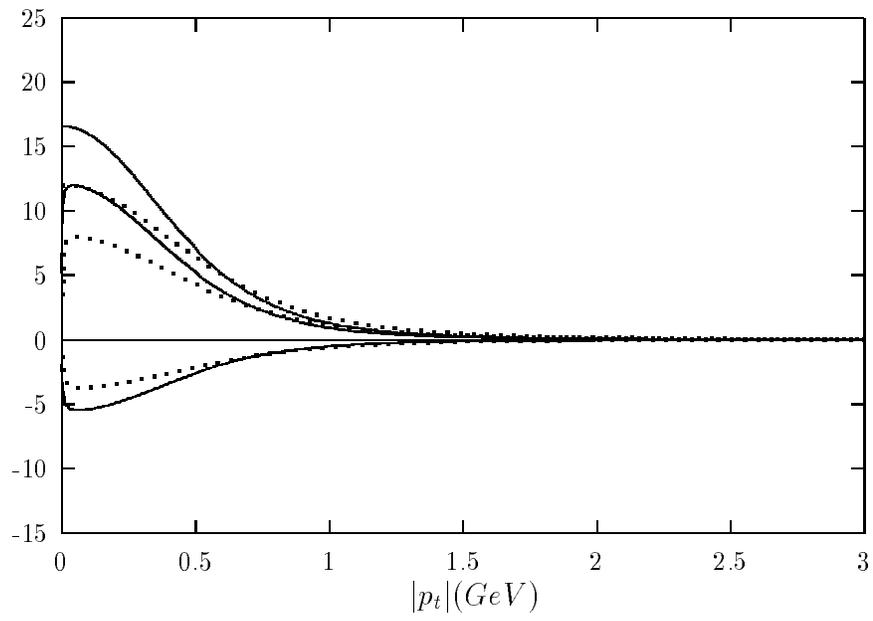
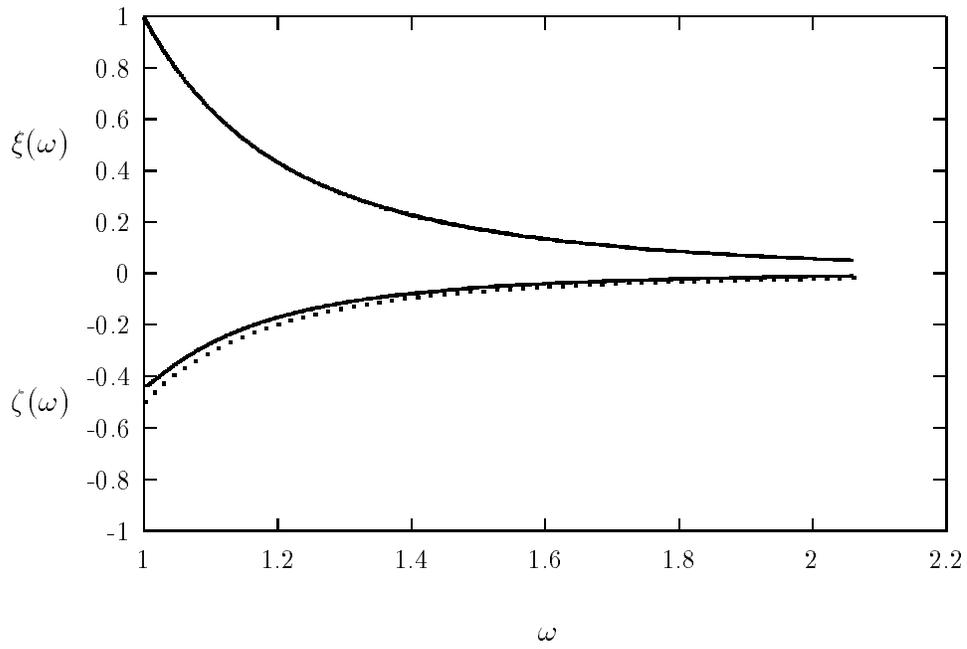
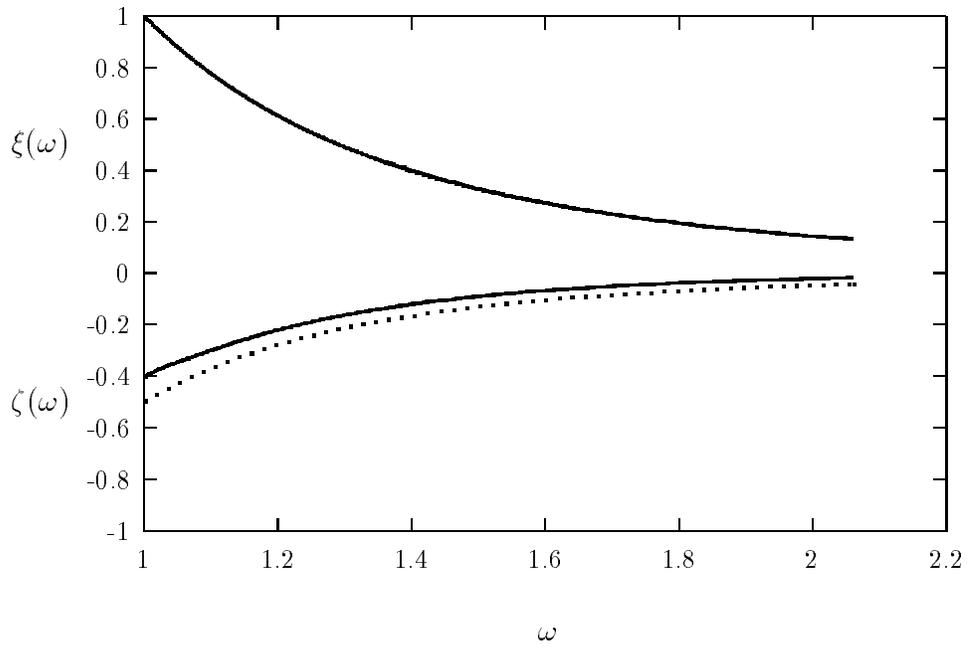
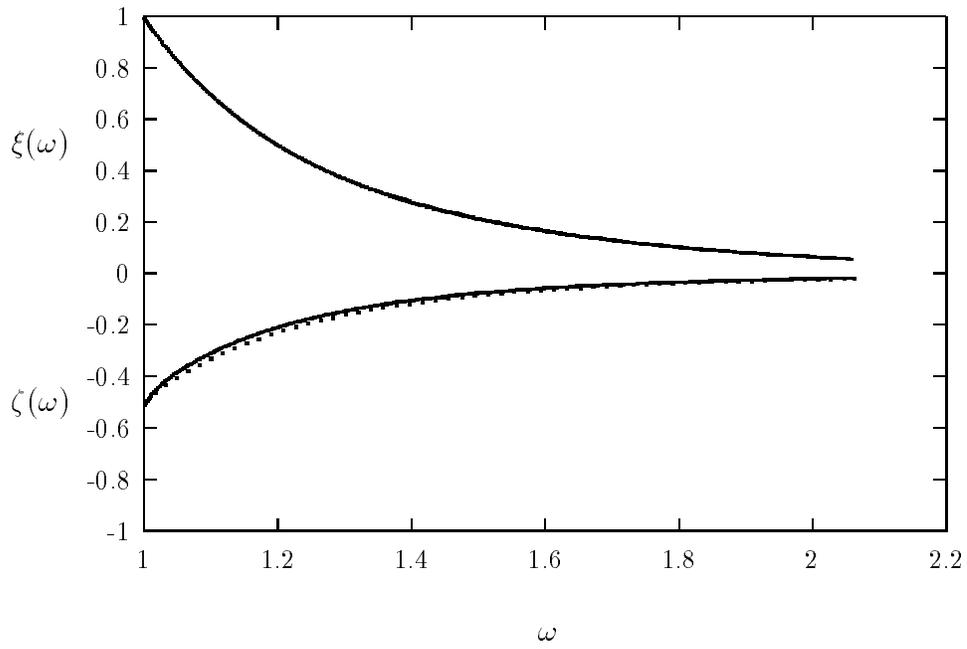
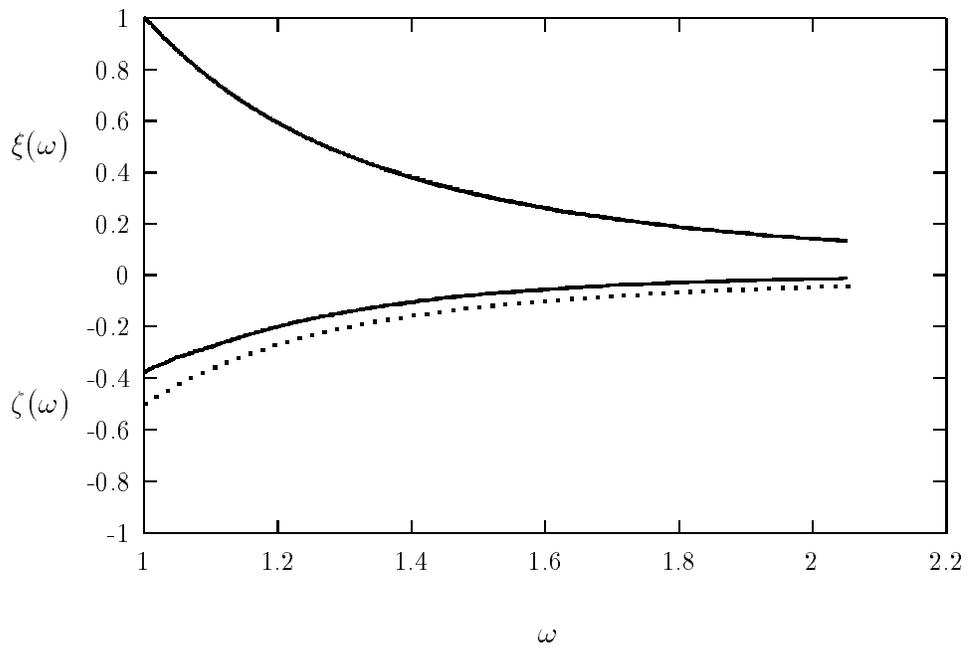
\end{document}